\documentclass{article}
\usepackage[utf8]{inputenc}
\usepackage[T1]{fontenc}
\usepackage{lmodern}
\usepackage{hyperref}
\usepackage{graphics}
\usepackage{wrapfig}
\usepackage{graphicx}
\usepackage{latexsym}
\usepackage{amsmath}
\usepackage{amssymb}
\usepackage{authblk}
\usepackage{pgf}
\usepackage{wrapfig}
\usepackage{pgf,pgfarrows,pgfnodes}
\usepackage{colortbl}
\usepackage{xcolor}
\usepackage{tcolorbox}
\usepackage{makeidx}
\usepackage{textcomp}
\usepackage{multicol}
\usepackage{multirow}
\usepackage{enumerate}
\usepackage{subfigure}
\usepackage{array}
\usepackage{geometry,amsthm,latexsym}
\usepackage[all,2cell]{xy} \UseAllTwocells \SilentMatrices
\usepackage{ragged2e}
\usepackage{natbib}
\bibpunct{(}{)}{;}{a}{}{,}
\newcommand{\apj}{ApJ}
\newcommand{\apjs}{ApJS}
\newcommand{\aap}{A\&A}

\newcommand{\aj}{AJ}
\newcommand{\apjl}{ApJ}

\newcommand{\araa}{ARA\&A}
\newcommand{\nat}{Nature}
\newcommand{\mnras}{MNRAS}

\newcommand{\pasj}{PASJ}

\usepackage[nottoc]{tocbibind}
\definecolor{c1}{rgb}{0,0,0.5}
\hypersetup{
colorlinks,%
citecolor=c1,%
filecolor=c1,%
linkcolor=c1,%
urlcolor=blue
}

\graphicspath{{./}{Figuras/}}  

\usepackage{graphicx}
\usepackage{txfonts}

\title{Global kinematics study of OH masers in W49N}

\author[1]{Mendoza-Torres, J.E.}
\author[1]{Ju\'arez-Gama M.}
\author[2]{Rodr\'iguez-Esnard, I.T.}

\affil[1]{Instituto Nacional de Astrof\'isica, \'Optica y Electr\'onica, C. Luis Enrique Erro No. 1, Tonantzintla, Pue. CP 72840, M\'exico, e-mail: \url{mend@inaoep.mx} }
\affil[2]{Instituto de Geof\'isica y Astronom\'ia, Street 212, No. 2906, between 29 and 31. CP 11600, La Coronela, La Lisa, La Habana, Cuba}

\date{}

\begin{document} 
\maketitle
 
\begin{abstract}
Star formation is underway in the W49N molecular cloud (MC) at a  high level of efficiency, with almost twenty ultra-compact (UC) HII regions observed thus far, indicating a recent formation 
of massive stars. 
Previous works have suggested that this 
cloud is undergoing a global contraction. 

We analyse the data on OH masers in the molecular cloud W49N, 
observed with the VLBA at the 1612, 1665, and 1667 MHz transitions in left circular polarization (LCP) 
and right circular polarization (RCP) with an aim to study the global kinematics of the masers.

We carried out our study based on the locations and observed velocities of the maser spots, V$_{obs}$.
We found the location ($\alpha, \delta$)$_{m}$ of the maximum correlation between V=V$_{obs}$-V$_{sys}$ (with V$_{sys}$ the systemic velocity) and distance to it. 
The velocities were fitted to the straight line of V$_{obs}$-V$_{sys}$ versus d$_{(\alpha, \delta)m}$, resulting in V$_{ftd}$.
The difference between the fitted values and those obtained from observations is $\Delta $V=(V$_{obs}$-V$_{sys}$)-V$_{ftd}$.
The V$_{obs}$-V$_{sys}$ velocity shows a gradient as a function of the distance to ($\alpha, \delta$)$_{m}$, where the closer spots have the largest velocities.
Spots with similar velocities are located in different sectors, with respect to ($\alpha, \delta$)$_{m}$. 
Then, we assumed that the spots are moving towards a contraction centre (CC$_{OH}$), which is at the apex of a CONUS. 
We also assumed that the distance of each spot to CC$_{OH}$ is d$_{cc}$ =$\sqrt{2}$~d$_{(\alpha, \delta)m}$ and that they fall with a velocity V$_{CC}$=$\sqrt{2}$V$_{ftd}$, with the total velocity being  V$_{Tot}$=V$_{CC}$+$\sqrt{2}$ $\Delta $V.
Using this velocity, we estimated the free-fall velocity.

The coordinates of $(\alpha, \delta)_m$ are effectively ($\alpha_{2000}$=19:10:13.1253, $\delta_{2000}$=9:6:13.570).
The observed dispersion with respect to the global trend against $d_{cc}$, shows a maximum at 0.12 pc, with a decay from 0.12 to 0.19 pc, which is faster
than that taking place between 0.19 and 0.42 pc.
Based on $V_{tot}$, an inner mass of M$_{inn}$=2500 $M_{\odot}$
was estimated.
In addition, the estimated accretion rate is 
$\dot M=$1.4$\times$10$^{-3}$ M$_{\odot}$yr$^{-1}$, 
which requires a time of t$_{inn}$=1.8$\times$10$^6$ yr 
to accumulate M$_{inn}$.
The free-fall time, assuming n=1$\times$10$^{-4}$ cm$^{-3}$,
is t$_{ff}$=3.4$\times$10$^5$ yr.
Performing the same procedure with published data that are of lower
spatial resolution (than the VLBA data) produces similar results. 
For example, based on the available data, 
we find that
$(\alpha,\delta)_m$ = (19:10:13.1392, 9:6:13.4387) J2000, which is 
at $\lesssim$ 0.3 asec from what has been calculated with the VLBA data, with an estimated inner mass of 2700 M$_{\odot}$.
A sub-collapse appears to be taking place in the region traced by the OH maser spots.
Based on methanol maser cloudlets data, which lie in a smaller region, another possible centre of contraction is identified, which could be due to a sub-collapse towards a 75 M$_{\odot}$ inner mass.
The velocities of the OH spots at W49N, along with their 
positions with respect to $(\alpha, \delta)_m$, make it possible 
to trace a global kinematics that is apparently due to a sub-collapse in the W49N MC.
\end{abstract}

\section{Introduction}
The W49N molecular cloud (MC) is part of the W49A giant molecular 
cloud, located at a distance of 11.1 kpc \citep{Zhang2013}.
The radial velocity of W49N has been estimated to be 
approximately 8.0 $km~s^{-1}$ \citep{Welch87}.  
In this region, about 20 HII regions have been observed 
\citep{Dreher84, DePree00, DePree04} 
in different spectral lines \citep{DePree97, Kulczak17}
and in the radio continuum \citep{Wilner01, Galvan13}.
The HII regions, identified by \citet{Dreher84} as 
ultra-compact (UC), are thought to be driven by O stars
\citep{Wilner01},
making this region one of the sites at our Galaxy characterised by the 
active formation of massive stars, with some of them estimated 
to be M$_{*}\simeq$25-35 M$_{\odot}$ and even 
M$_{*}\simeq$45~M$_{\odot}$ \citep{Smith09}.

Masers of OH \citep{Johnston82, Kent82, Argon00} and of 
H$_2$O \citep{Gwinn89, Gwinn92, Gwinn94, Takefuji16} have been observed in W49N.
The OH spots detected by \citet{Kent82} with 
the VLBI have velocities ranging from 4.6 to 21.1 $km~s^{-1}$ 
and \citet{Argon00} those registered with the VLA, OH spots 
have velocities between 8 and 22 $km~s^{-1}$. 
The spots with more red-shifted velocities are mainly
located at the west (W), while the less red-shifted appear 
at the east (E).
H$_2$O maser spots observed by \citet{Gwinn92} at W49N, have been interpreted as arising at a bipolar outflow of a newly formed star.
The spot velocity increases with distance, to a common centre of the flows (denoted by a square in Figure~\ref{FmapOHyHII}), up to velocities greater than 200 $km~s^{-1}$ \citep{Gwinn92, Gwinn94}.
Each outflow extends about 1 arcsec from their common centre, denoted by a square in Figure~\ref{FmapOHyHII}.

At other sources, the proper motions of OH maser spots 
have allowed for the local kinematics to be established as likely being due to the expansion of HII regions 
\citep{Bloemhof92, Fish07, Liu15}, 
as well as jets \citep{Argon03}, outflows \citep{Fish11},
and shells and rings \citep{DePree00}.

In W49N, a velocity gradient has been identified, based on a 
set of HII regions that form a ring \citep{Welch87}
and it has been suggested that the cloud is in contraction. 
On the other hand, \citet{Serabyn93} suggested that the kinematics is due to the collision of two 
molecular clouds. 
\citet{Rudolph90} found inverted P-Cygni profiles in 
the HCO$^+$ molecule in W51, which is an association of almost a 
dozen ultra-compact HII (UC-HII) regions and estimated 
that it is engaged in a global collapse. 
They get a mass accretion rate of $\dot M$=10$^{-4}$ $M_{\odot}$ yr$^{-1}$ and an inner mass of 10$^5$ M$_{\odot}$.
 
Methanol masers have been seen at W49N \citep{Bartkiewicz14, Pandian11}.
They are found to be associated with high-mass star formation \citep[e.g.][]{Sanna10a, Moscadelli11} and are capable of tracing the kinematics close to the young stellar object (YSO, \citet{Moscadelli11}).
At G16.59–0.05, methanol maser cloudlets trace a rotation around a central mass of about 35 solar mass \citep{Sanna10a}.

A velocity gradient of each methanol maser cloudlet, V$_{grad}$, has been seen in a number of sources.
Here, the gradient at the cloudlets is compared with the velocity gradient seen at the radial velocity of the cloudlets.
From multi-epoch observations, it is found that the velocity gradients and the proper motion vectors point in the same direction in the sky \citep{Moscadelli11}.
 
There are still unsolved problems related to the formation of 
high-mass stars \citep[see][]{Motte18, Zinnecker07}. 
For example, to explain how they can 
grow to high masses, even under the action of 
radiation feedback, and how many of them can grow in a 
given region of a molecular cloud.  
The latter is known as mass 
segregation, in hierarchical substructures, the most massive 
protostellar objects are  generally located in the centre of a cluster surrounded by objects of lower mass \citep{Bonnell03}.
One possibility for having many massive stars 
is that the cores are massive enough when the star formation process begins. 
The switch-off of the fragmentation process can 
allow for the conservation of massive cores, 
which could be due to a slow build-up supported by turbulence 
or thanks to the stabilisation of massive clusters for a 
magnetic field \citep[see][]{McKee02, Commercon11, Myers13}
Also, as suggested by \citet{Welch85}, 
the accretion on a global scale, lasting for 
long timescales, could lead 
to the accumulation of high masses. 

The models applied by \citet{Welch85} and
\citet{Rudolph90} to interpret the velocity field, as well as 
the observed masses, agree with the idea that the OB 
associations are formed by the dynamic collapse of large 
regions in molecular clouds.
A better understanding of the kinematics of the MC W49N will help us
to gain a better insight into the processes of star formation 
and, in particular, the formation of OB associations.
  
\section{Data analysis and results}
The data were obtained in the OH lines at the following frequencies: 1667.35903, 1665.4018, and 1612.23101 MHz, hereafter 
referred to as 1667, 1665, and 1612 MHz, respectively.  
We are using the same data as \citet{Deshpande13} with no new calibrations. The observations
were made on 6 October 2005, with the Very Long Baseline Array (VLBA) interferometer. 
Left circular polarization (LCP) and right circular polarization (RCP) are observed simultaneously in 
about 240 spectral channels with a resolution of 
0.1 $km~s^{-1}$ in a range of about 22 $km~s^{-1}$. 
The beam size is about 20 milliarcseconds (mas) by 
15 mas at a position angle of 84 degrees.
The reference coordinates are $\alpha$=19:10:13.2091 
and $\delta$ = 09:06:12.485 (J2000).

\begin{figure}[ht!]
\centering 
\includegraphics[height=9.7 cm,width=8cm]{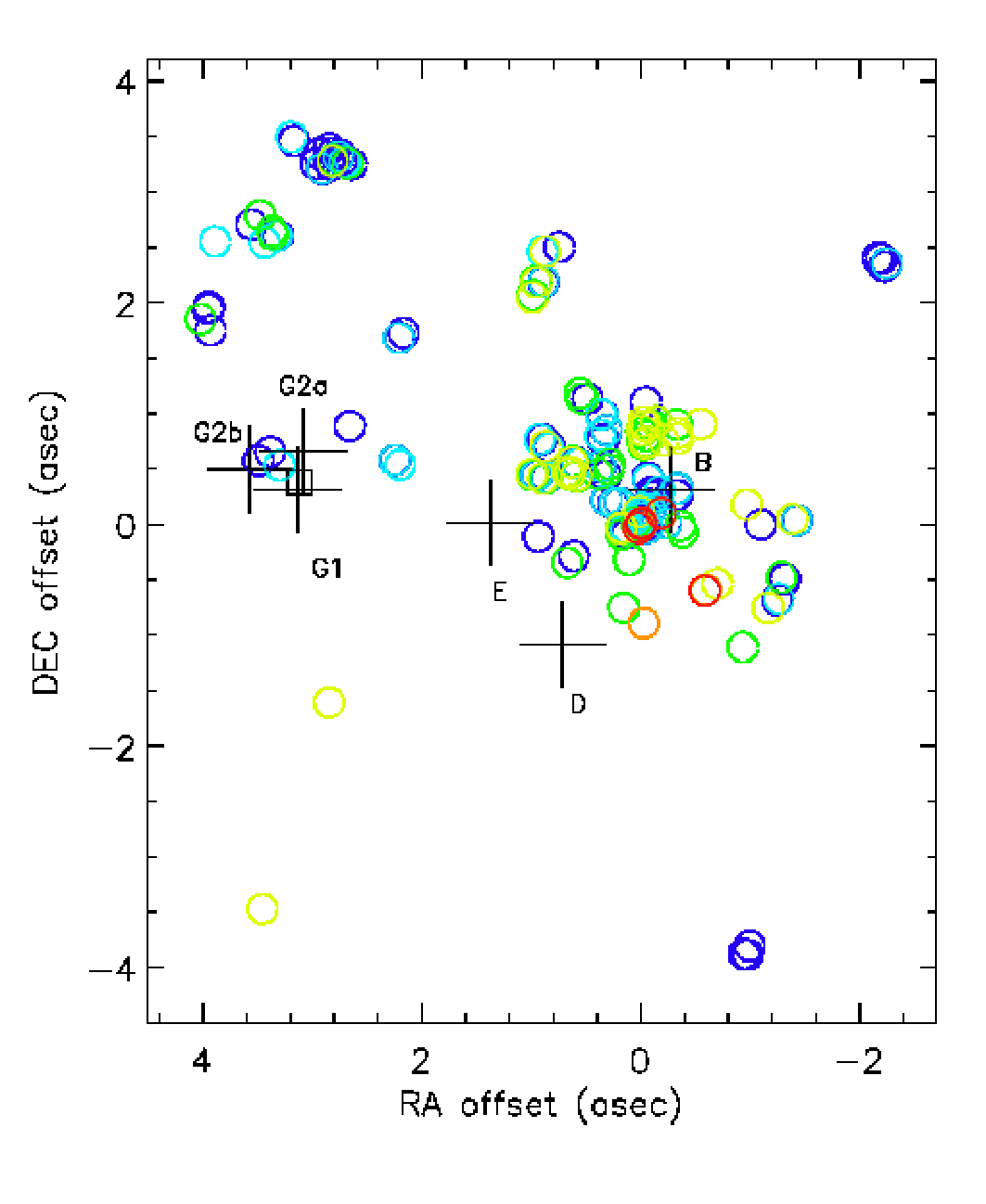}
\caption{Map of the OH maser spots respect the reference coordinates. 
The (0,0) location corresponds to $\alpha$=19:10:13.2091, $\delta$=09:06:12.485 (J2000).
Respectively, the dark and light blue circles represent the 1667 MHz LCP and RCP
maser spots; green and yellow is for 1665 MHz LCP and RCP;
and orange and red circles for 1612 MHz LCP and RCP.
The crosses indicate the locations of HII regions, reported by
\citet{DePree97}, with the corresponding letters used to
refer to each of them.
The square that overplots to the cross, which represents G1, corresponds to
the location of the centre of the bipolar outflow of \citet{Gwinn92}.}
\label{FmapOHyHII}
\end{figure}

\begin{table}[ht!]
\begin{center}
\caption{Number of spots recorded at each frequency and polarization.} 
\begin{tabular}{|l||c|c|c|} 
\hline 
Freq & LCP & RCP & L+R \\ 
\hline
1667 &  73 &    44 &    117 \\ 
1665 &  42 &    35 &    77 \\ 
1612 &  6 &     5 &     11 \\ 
\hline 
\end{tabular}
\label{TPWVobsYftd00y12}
\end{center}
\end{table}

To identify maser spots and
estimate their flux, central velocity, and coordinates, 
Astronomical Image Processing System (AIPS) routines were used. 
In total, 205 spots were identified, in the three 
observation frequencies at the two polarizations 
(Table 1 and Figure~\ref{FmapOHyHII}). 
The size of all the spots is greater than the interferometer beam, indicating that the spots are spatially resolved. 
The field of the OH spots is at the SW end of the Welch ring of HII regions, 
in a field made up of about 6 asec  sides (Figure~\ref{FmapOHyHII}). 
The observed velocity of the spots, V$_{obs}$, is taken respect to the systemic
velocity of 8.0 $km~s^{-1}$ \citep{Welch87} and it spans about 18.6 $km~s^{-1}$. 
The spots with red-shifted V$_{obs}$ are predominantly 
located in the W of the map, while the blue-shifted in the E,
as reported by \citet{Kent82}. 
This is particularly clear for the frequencies where more spots
are seen (1665 and 1667 MHz) at both polarizations (LCP and RCP).

The bipolar outflow lies on the HII region denoted as G1 by \citet{DePree97}, as also found by \citet{Bloemhof00}.
The common centre of the outflows, traced by the H$_2$O maser spots, becomes -- at a location given by an offset in mas of (3110, 380) -- relative to the VLBA reference coordinates.

\begin{figure}[ht!]
\centering
\includegraphics[height=6.3cm,width=6.6cm]{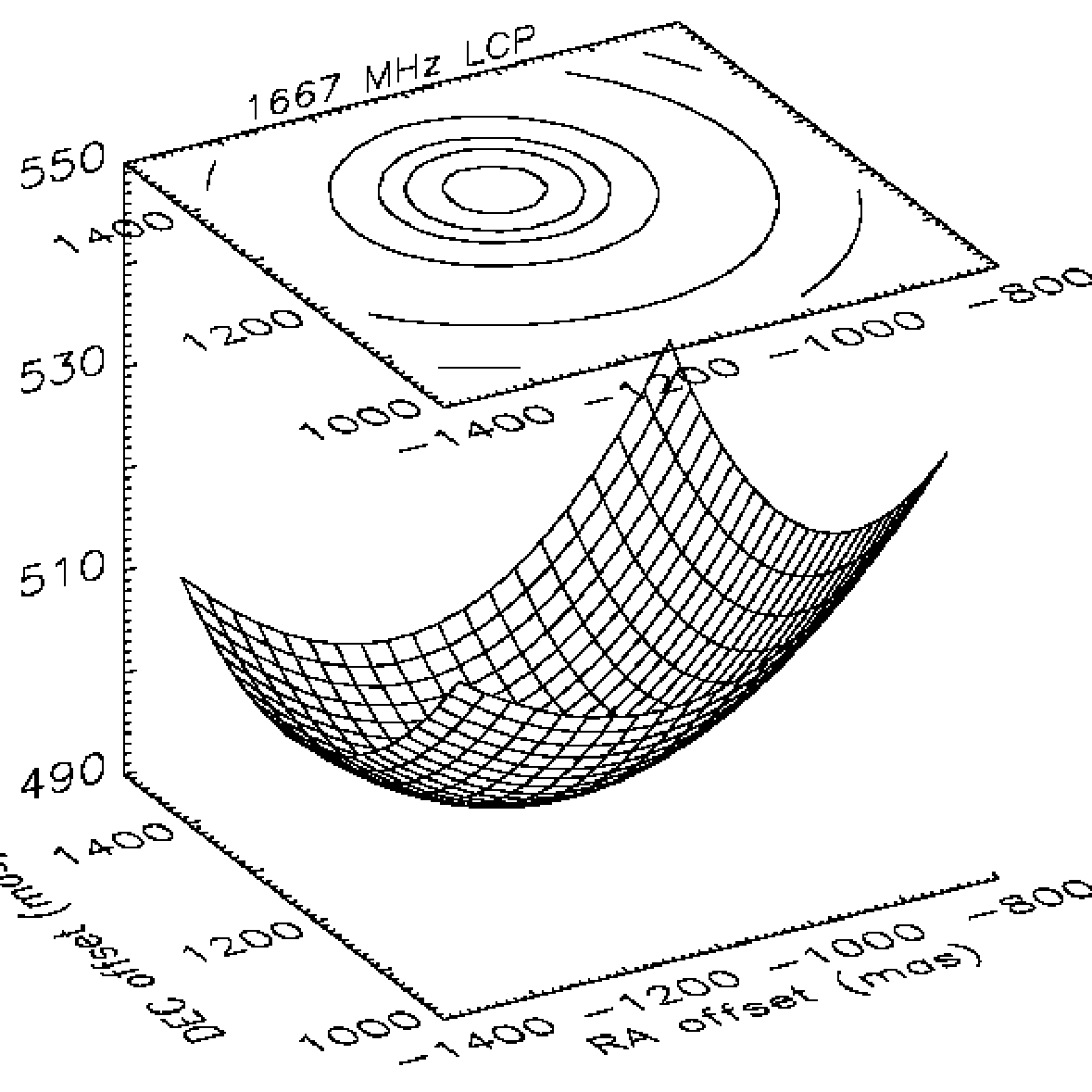}
\includegraphics[height=6.3cm,width=6.6cm]{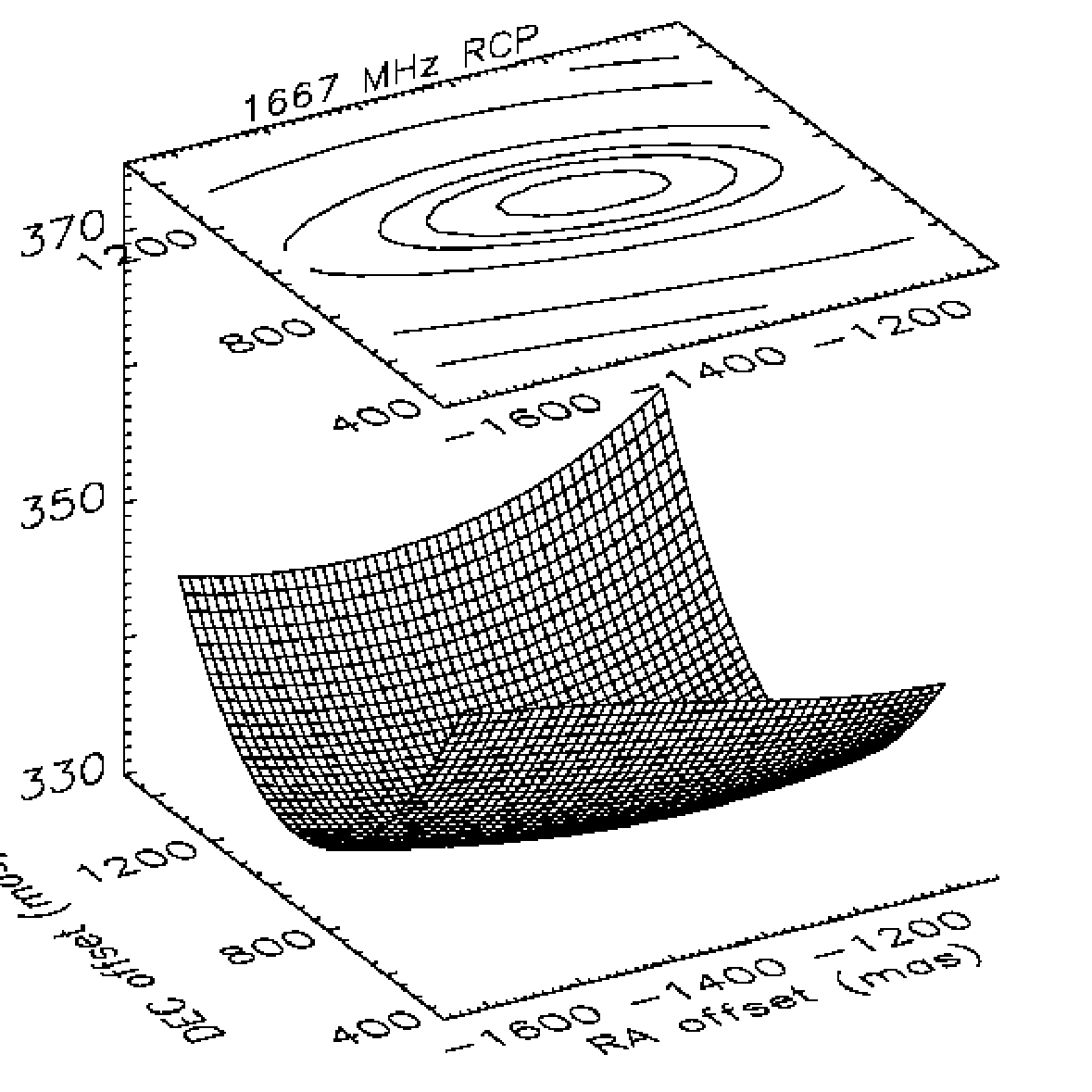}
\includegraphics[height=6.3cm,width=6.6cm]{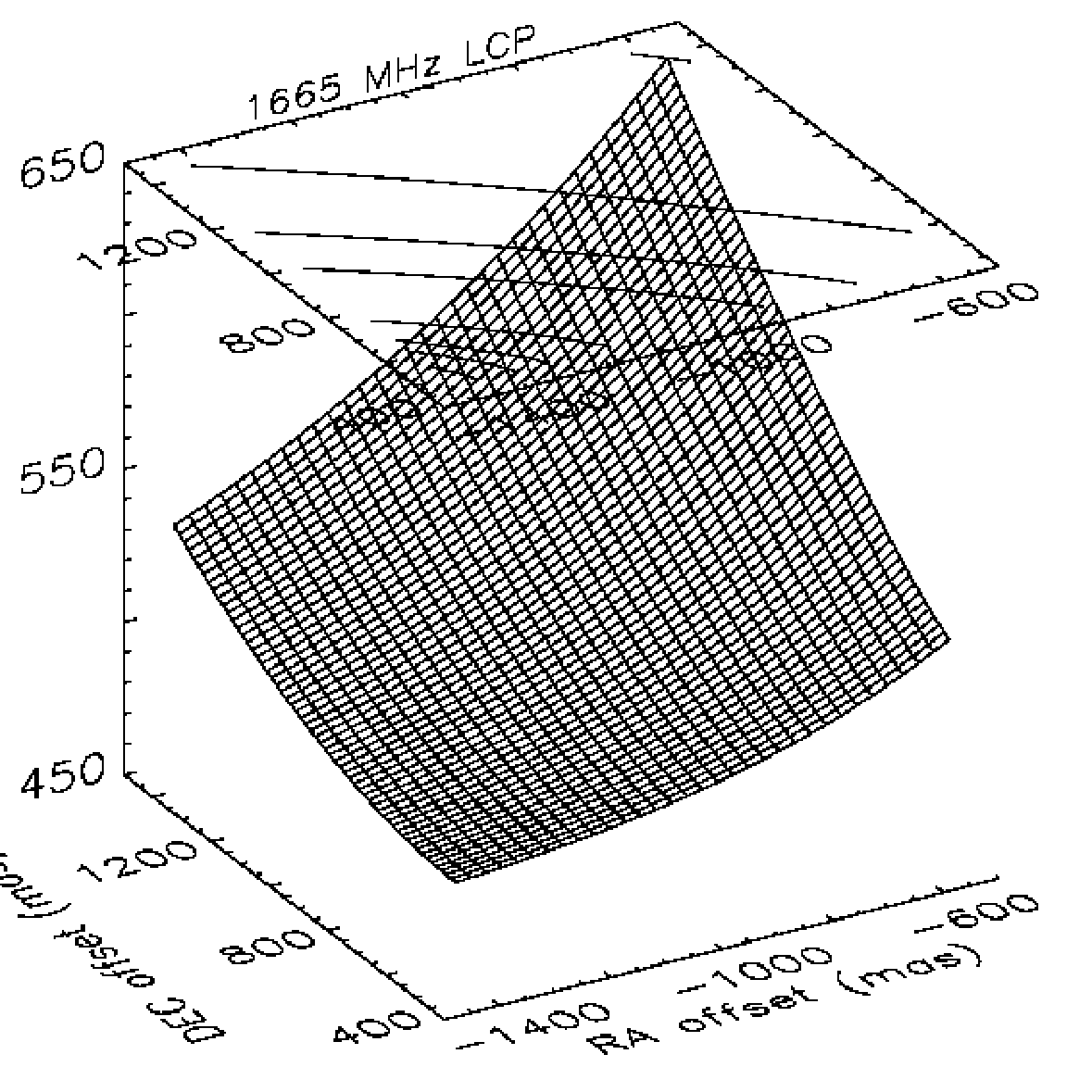}
\includegraphics[height=6.3cm,width=6.6cm]{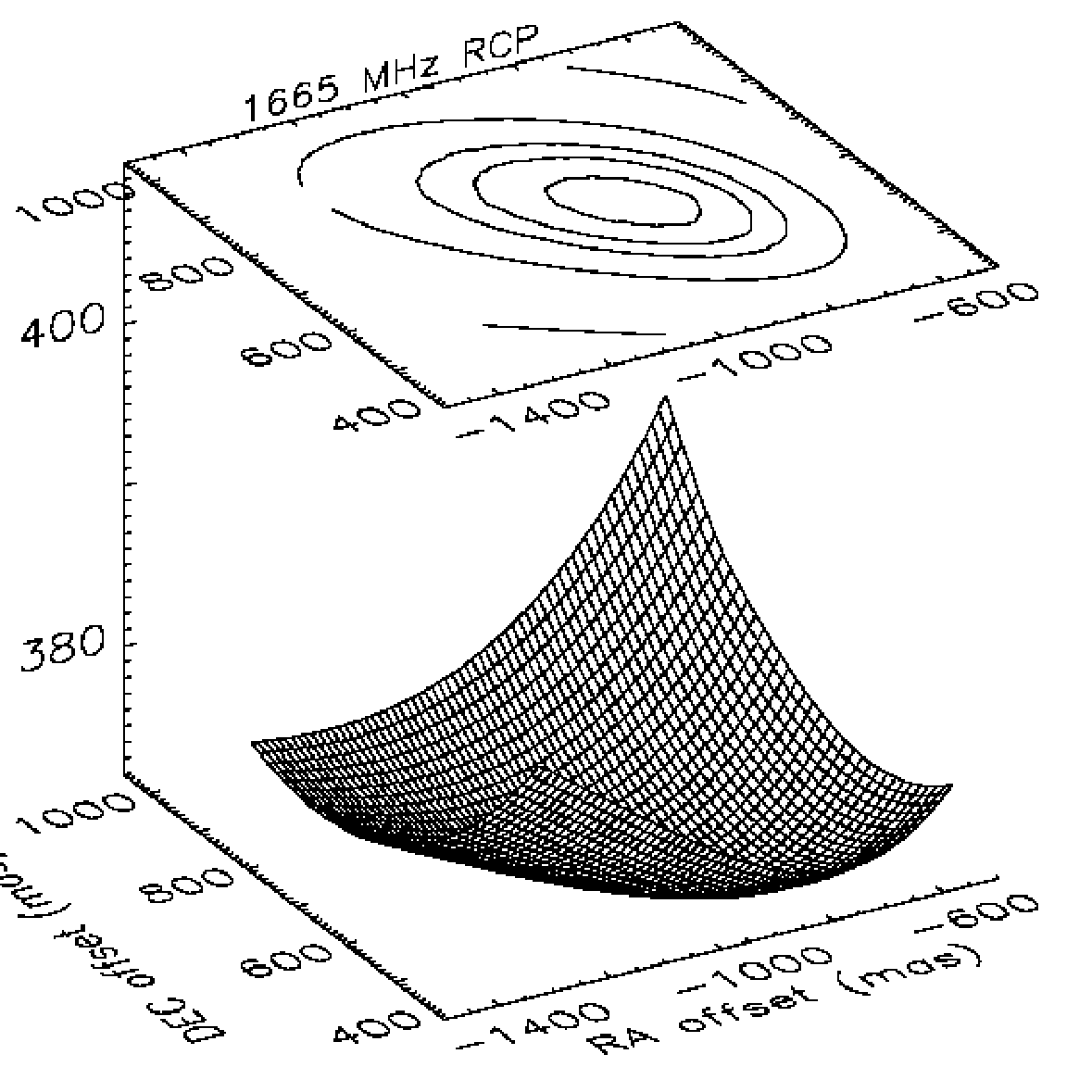}
\caption{$\chi^2$ values, in contour 
levels and 3D surfaces around the minimum values 
for 1667 MHz LCP (top), 1667 MHz RCP (from top to bottom),
and 1665 MHz LCP and 1665 MHz RCP (bottom).}
\label{FsurfCHi2}
\end{figure}

In order to study the global kinematics of the OH spots, 
we tested the global velocity field to see if it shows a gradient. 
In spots separated by distances $<$ 1 asec, gradients 
are observed, however, we did not test single groups of spots,
but all the spots across the whole field. 

As  mentioned above, the OH maser spots that are more red-shifted are in the east, while the less red-shifted ones are in the west.
We searched for the possibility that the velocity of the spots
versus the distance, with regard to a given location between the groups 
with different shifts, fits a straight line function.
With this purpose, we applied the $\chi^{2}$ test.
The $\chi^{2}$ error is the sum of squared differences between the
values of V$_{obs}$-V$_{sys}$ and those of V$_{ftd}$, where these last
values result from the fit for the distances, $d_{ij}$, of the spots to the location
($\alpha_i$, $\delta_j$), in mass offsets respect the VLBA reference.
The smaller the $\chi^{2}$ value, the better the fit.

We estimated $\chi^{2}$ for different locations ($\alpha_i$, $\delta_j$), 
regularly separated in both, $\alpha$ and $\delta$, 
thus forming a mesh.
First, we used a mesh that encompasses the whole field of view (FOV) and
for each point of the mesh, the test was applied.
Based on these results, it is seen that, certainly, 
in the middle locations of the FOV, $\chi^{2}$ takes lower
values than the rest locations of the FOV.
Then, another mesh, with regularly spaced points by 
$\Delta\alpha$ = 10 mas offsets in $\alpha$ and $\delta$
(i.e. with respect to the reference coordinates) and in
the range, in mas offsets, from 5000 to -4000 in $\alpha$ 
and from -4500 to 4500 mas in $\delta$ is used 
to identify the minimum value of $\chi^{2}$. 
For each set of $d_{ij}$ distances, the correlation 
between V$_{obs}$ and $d_{ij}$ was also computed.

We calculated the coefficients of the straight line fitted to 
V$_{obs}$-V$_{sys}$ against $d_{ij}$ (we remember that for each 
point of the mesh, we have a series of distances $d_{ij}$), 
obtaining fitted values, V$_{ftd}$.
Since V$_{sys}$ is here considered a constant (V$_{sys}$=8.0 $km~s^{-1}$), then the functional relation between V$_{obs}$-V$_{sys}$ and d$_{(\alpha, \delta)m}$ (seen in Figure 3) is attributed to the value of V$_{obs}$.
Furthermore, we could refer to this functional relation, with the aim to abbreviate the descriptions, by just mentioning V$_{obs}$ and not the whole expression of V$_{obs}$-V$_{sys}$. 
The difference between the fitted and the observed velocity $\Delta $V=(V$_{obs}$-V$_{sys}$)-V$_{ftd}$ was computed as well as the standard deviation (SD) of $\Delta $V
for each testing point.

In Figure~\ref{FsurfCHi2}, the $\chi^2$ values are shown in contour 
levels and 3D surfaces, around the minimum value. 
As can be seen from the plots of Figure~\ref{FsurfCHi2}, 
there is a clear minimum of $\chi^2$ at 1667 LCP and 
RCP and at 1665 RCP. 
This means that there is a best fit for a given location for these
frequencies and polarizations among all the
$(\alpha_i, \delta_j)$ values of the mesh.
For 1665 LCP, there was not a single minimum of $\chi^2$.
However, as can be seen from Figure~\ref{FsurfCHi2}, the $\chi^2$ 
values are lower for the zone of the minima at 1667 LCP
and RCP, as well as at 1665 RCP. 
In this zone, the value of correlation (which is the highest at 1665 L) 
repeats for a number of locations.
The same happens for the minimum of $\chi^2$ and the minimum of SD.
Then, averaging the coordinates of the highest correlations, 
we estimated a location that is considered the site of
the maximum. 
For 1612 LCP and 1612 RCP, 
we did not make the above calculations due to the 
reduced number of spots (Table 1).

Based on the results of the above analysis, we 
identified the maximum correlation between 
V$_{obs}$ and $d_{ij}$, the minimum of $\chi^2$ 
and the minimum of the SD.
The maximum correlations, for 1667 LCP and RCP and for 1665 
LCP and RCP are, respectively, 0.91, 0.87, 0.87, and 0.72. 
On the other hand, the values of the minimum SD are 2.6, 2.8, 2.8, and 3.6 $km~s^{-1}$ (giving them in the
same order as the correlations).
The maxima of the correlation coincide with the minima
of $\chi^2$ and SD.
Using the coordinates of the maximum correlation at 1667 LCP 
and RCP and 1665 LCP and RCP, we calculated the average coordinates, 
$(\alpha, \delta)_m$, weighted with the number of spots at 
each frequency and polarization, which turns out to be 
(-1236.6,1085.1), relative to the reference coordinates or 
$\alpha$=19:10:13.1253, $\delta$=9:6:13.570 in J2000.

\begin{figure}[ht!]
\centering
\includegraphics[height=6.3cm,width=6.6cm]{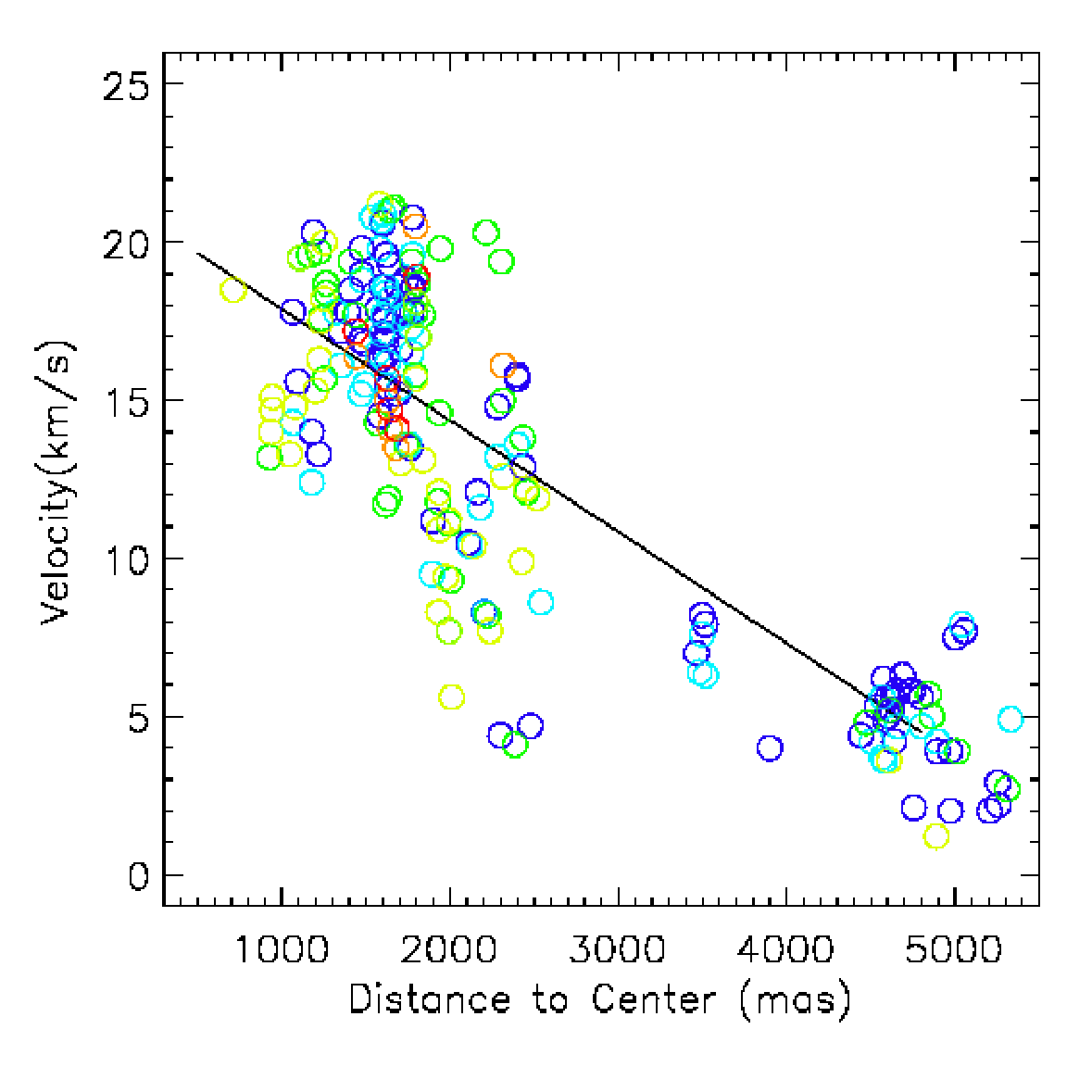}
\includegraphics[height=6.3cm,width=6.6cm]{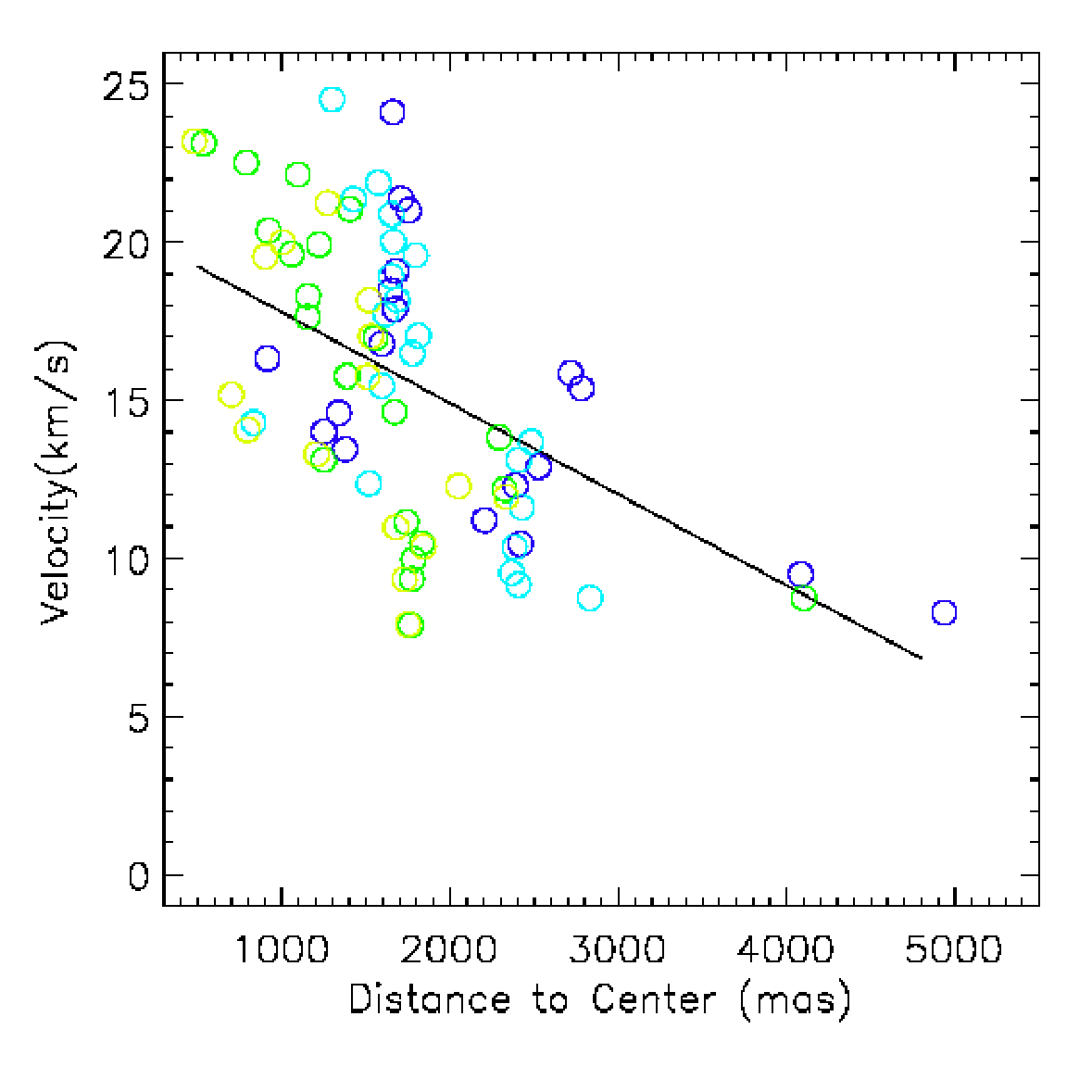}
\caption{ V$_{obs}$-V$_{sys}$ (with V$_{sys}$=8.0 $km~s^{-1}$) versus the distance to the $(\alpha, \delta)_m$ point for the OH VLBA data shown (top).
A clear trend is seen, where the velocity grows as the distance d$_{(\alpha, \delta)m}$ decreases. Same relation but for the VLA data of colour (bottom) taken from \citet{Argon00}. Symbols and colours are the same as in Figure~\ref{FmapOHyHII}.}
\label{FVelvsDistini}
\end{figure}
 
In the top panel of Figure~\ref{FVelvsDistini} a 
plot of V$_{obs}$-V$_{sys}$ (with V$_{sys}$=8.0 $km~s^{-1}$) versus d$_{(\alpha, \delta)m}$ 
for the VLBA data is shown.
A clear trend is seen, where the velocity grows as the
distance d$_{(\alpha, \delta)m}$ decreases. 
The velocities of the fitted line will be further referred to as V$_{ftd}$.
For comparison,  the \citet{Argon00} velocity
is also plotted with distance (bottom panel of Figure~\ref{FVelvsDistini}),
estimated to its reference 
coordinates (plotted in the same units as for VLBA data).
The coordinates $(\alpha, \delta)_m$, relative to the reference,
with respect to the OH VLBA observations,
are (-1029.9, 954.1), which is at a distance to the point 
estimated for the VLBA data that is less than 0.3 asec.

\begin{figure}[ht!]
\centering
\includegraphics[height=6.3cm,width=6.6cm]{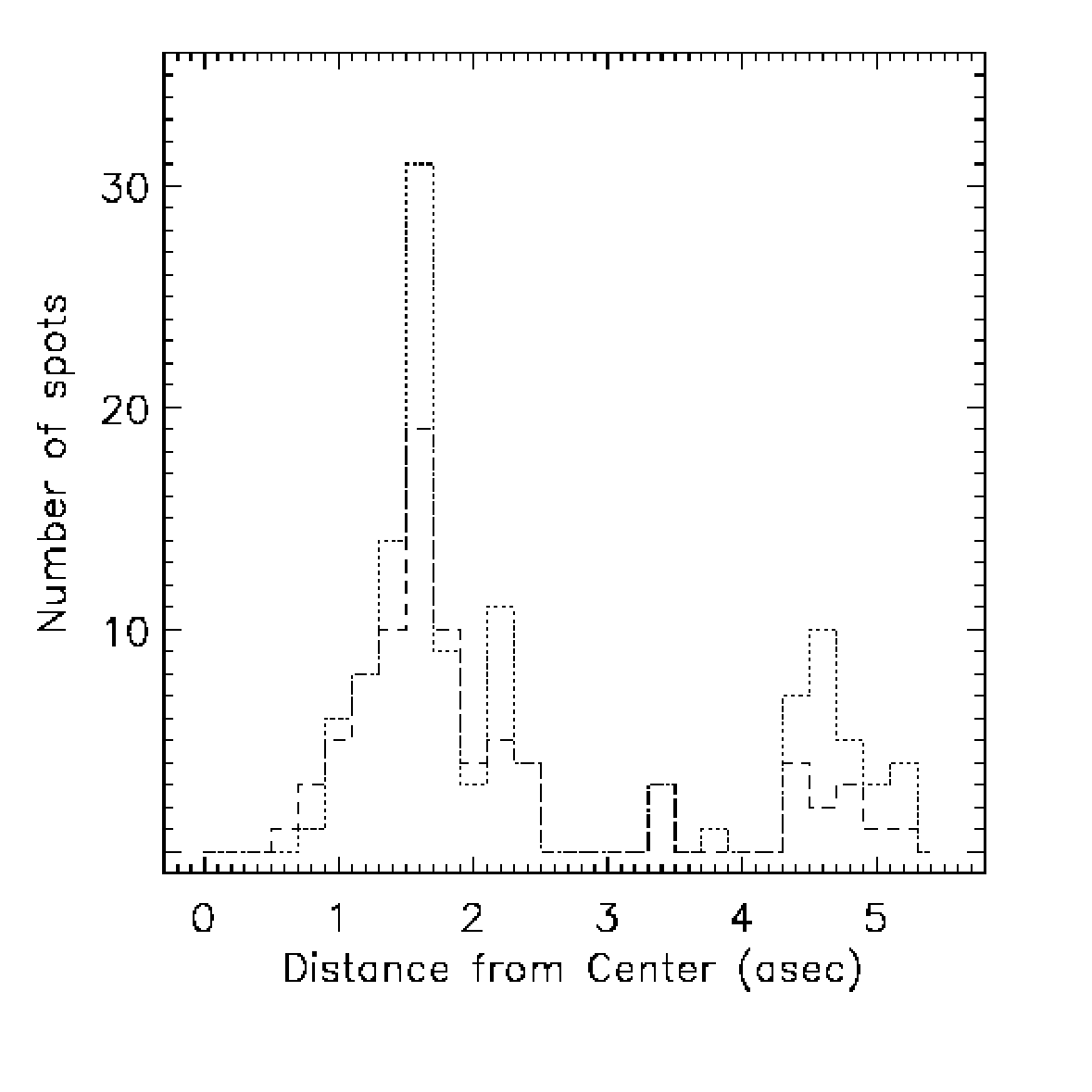}
\includegraphics[height=6.3cm,width=6.6cm]{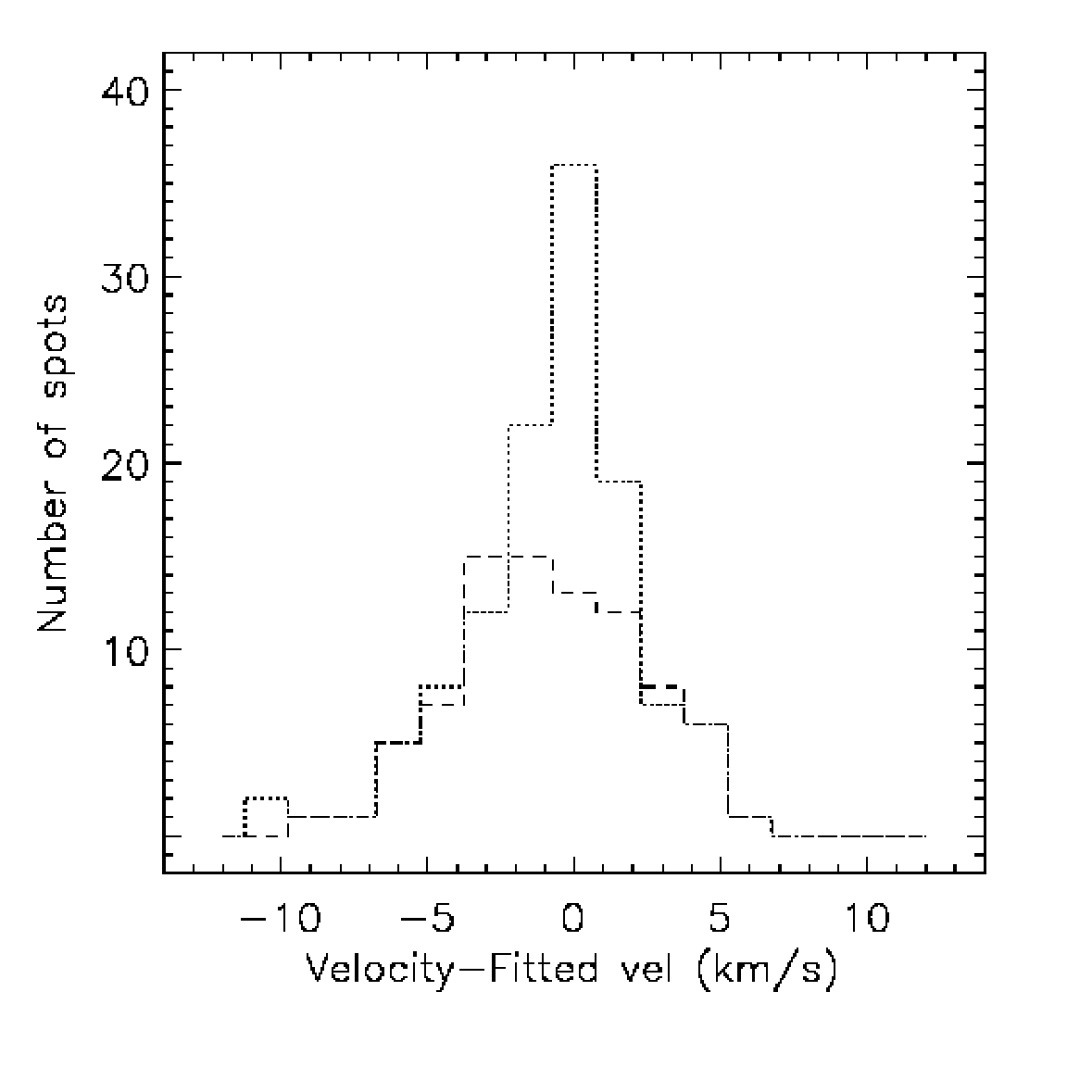}
\caption{Histogram of the number of spots (including the three frequencies) with the distance 
to $(\alpha, \delta)_m$, dotted lines are for LCP and dashed for RCP (top). 
{\it } Histogram of the number of spots with velocity (bottom). 
Lines are as in the top panel.}
\label{FhisDisHisVel}
\end{figure}

The distance histograms have been made by computing the number of spots of LCP and RCP, for a 
series of distance intervals (or bins, each of 0.2 asec width)
to $(\alpha, \delta)_m$, for the data of 
the three frequencies (top panel of Figure~\ref{FhisDisHisVel}). 
The velocity histogram is shown in the bottom panel of Figure~\ref{FhisDisHisVel}.
It may be seen that there is no shift, neither in
distance nor in velocity, between the LCP and RCP polarizations.
In the subsequent steps of the analysis, we included both of them together.

\begin{figure}[ht!]
\centering
\includegraphics[height=7.6cm,width=9.5cm]{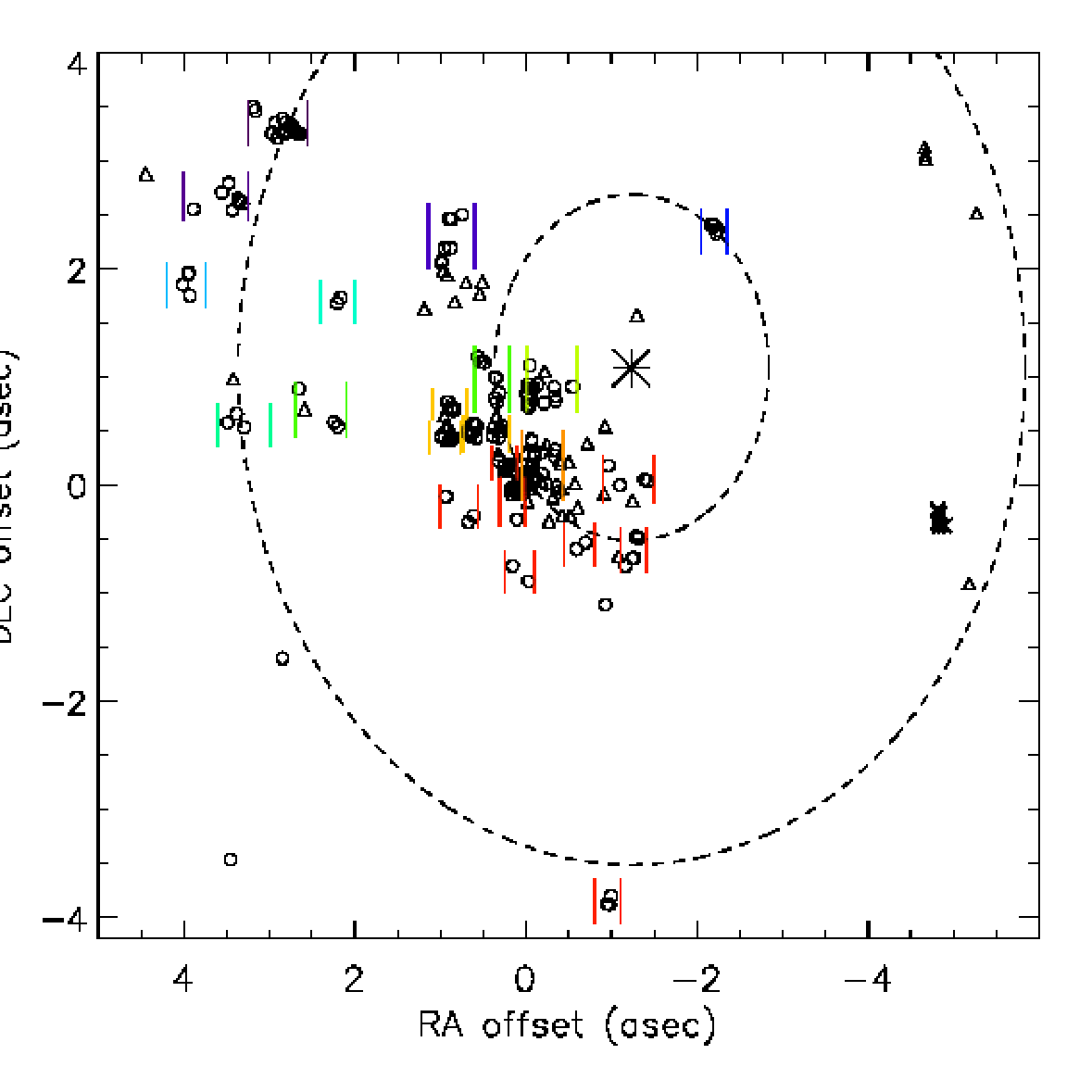}
\caption{Locations of the OH maser spots,
observed with the VLBA, are indicated with small circles.
These spots are the same as those plotted in Figure~\ref{FmapOHyHII} (also with circles).
The asterisk represents the $(\alpha, \delta)_m$ point. 
The groups used to estimate the average velocity
(Figure~\ref{FVelVSdistPCcc}) are indicated with vertical bars. The \citet{Argon00} OH maser
spots are indicated with triangles and the methanol maser cloudlets with crosses ($\times$).  
All the methanol maser cloudlets that lie in this FOV \citep{Bartkiewicz14, Pandian11} are on the west side, clumped together near the large dashed circle.}
\label{FCoordsROIs}
\end{figure}

In Figure~\ref{FCoordsROIs}, a map of the spots of all 
frequencies and polarizations is shown. 
The asterisk symbol represents the $(\alpha, \delta)_m$ point.
The small dashed circle indicates a distance of 1.6 asec 
to $(\alpha, \delta)_m$, which corresponds to the first
maximum of the histogram for distance to $(\alpha, \delta)_m$ 
for OH masers VLBA data. 
For the \citet{Argon00}  OH masers VLA data, the first maximum of the histogram also takes place at 1.6 asec.
The subsequent analysis made for VLBA data also leads to 
similar results with the \citet{Argon00} VLA data. 
The large dashed circle corresponds to a 
distance of 4.6 asec, where the other maximum of
the distance histogram (top panel of Figure~\ref{FhisDisHisVel}) takes place.
The spots in Figure~\ref{FCoordsROIs}, which are around the 
small circle, appear mostly in the east, but also spots are observed in the north-west  (NW) and the south (S), near this circle. 
The spots around the large circle also appear preferentially 
in the east and some in the south.
Taking the circles in Figure~\ref{FCoordsROIs} as a reference, 
it may be seen that the spots are located in various sectors and, 
without considering the dispersion of shorter spatial scales
($<$ 1 asec), the velocities of spots at 
equal distances to $(\alpha, \delta)_m$ are 
similar to each other.

\pagebreak

\section{Methanol maser cloudlets}
We compared the locations of the methanol masers, observed by \citet{Bartkiewicz14} and by 
\citet{Pandian11}, in W49N
with those of the OH maser spots, and their velocities are analysed to look how they fit in the model proposed for OH spots.
The \citet{Bartkiewicz14} observations were set to four pointing positions.
The coordinates (J2000) of the brightest spot at each position are given in their Table 2.
The masers detected at W49N are spread over an area of 1.3$^\prime\times$0.9$^\prime$.
In their Table B.1, the locations of the spots are given as angular offsets (in mas) with respect to the location of the brightest spot of each group.
The G43.165+00.013 group, which contains 12 spots, is the only one that lies in the FOV of the OH VLBA masers.
The coordinates of the brightest spot of this group are $\alpha$=19:10:12.882 and $\delta$=09:06:12.2299. 
Using the coordinates of this spot and the offsets of the other spots of the group (with respect to the former), we obtained their corresponding 
$\alpha$ and $\delta$ (J2000). 
Then, with these coordinates and the coordinates of the OH reference point ($\alpha$=19:10:13.2091, $\delta$=09:06:12.485), the spot coordinates were transformed to offsets, but now with respect to the OH VLBA reference.

\begin{figure}[ht!]
\centering
\includegraphics[height=5.3cm,width=5.6cm]{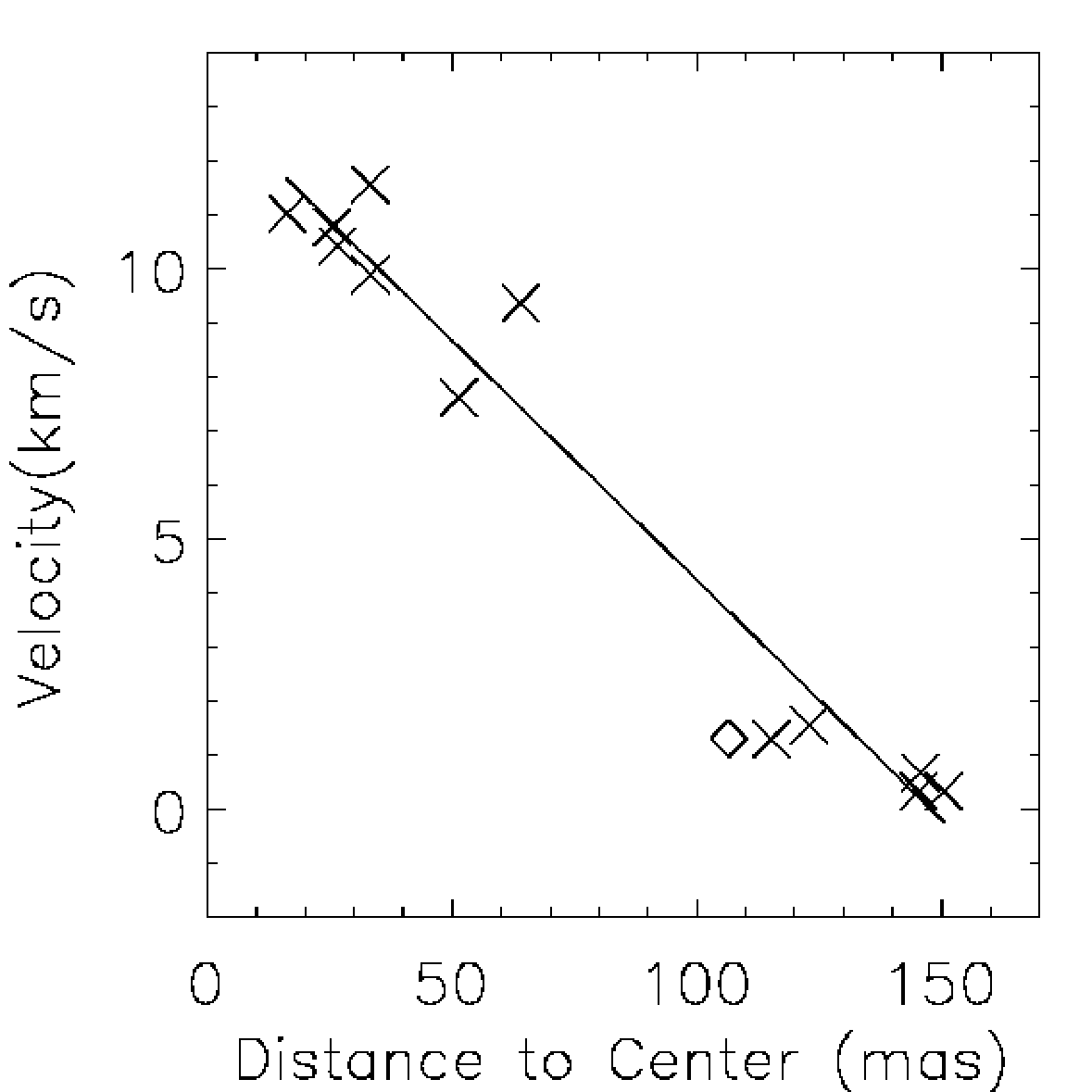}
\includegraphics[height=5.3cm,width=6cm]{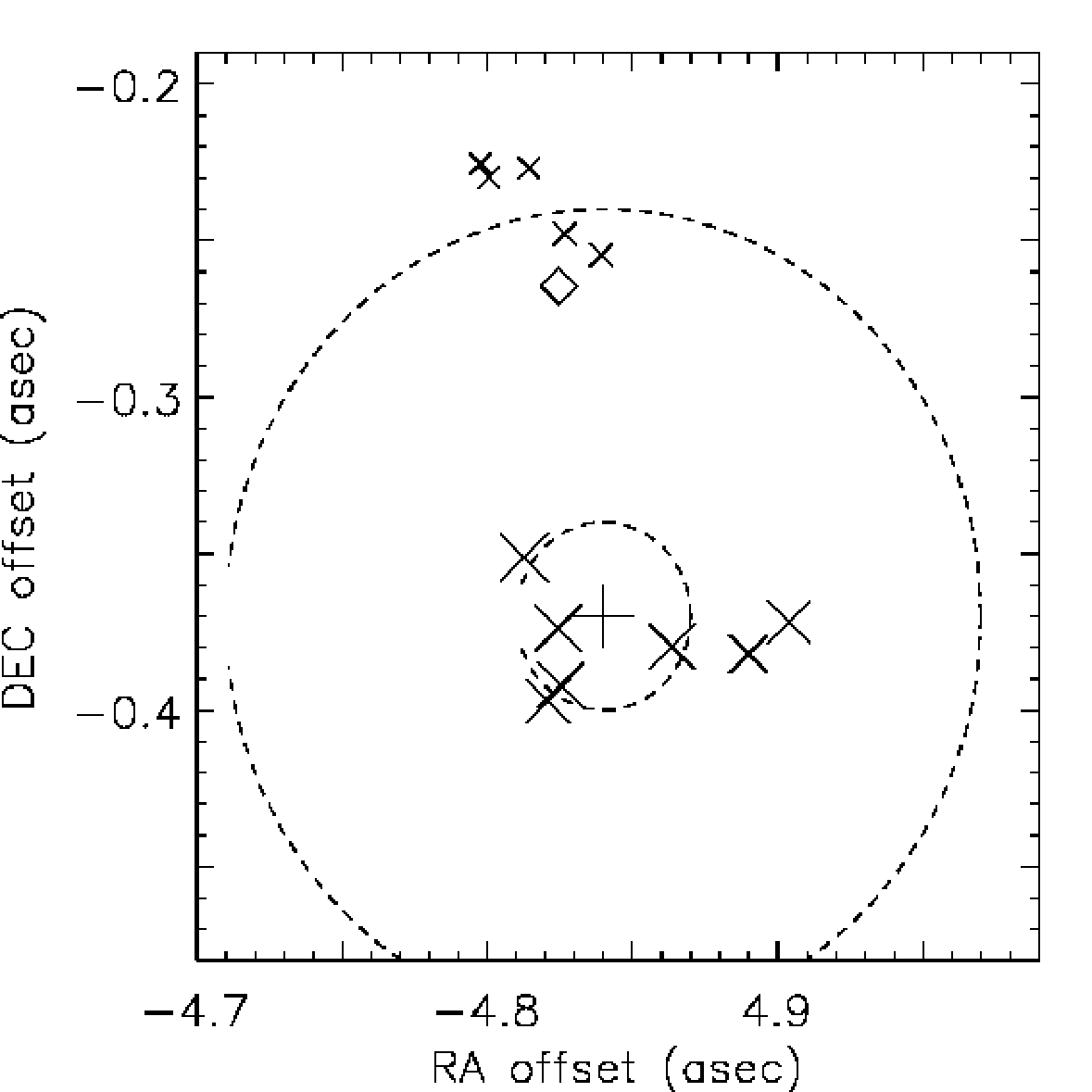}
\caption{{\it } Crosses indicate the V$_{pk}$-V$_{sys}$ velocities against the d$_{(\alpha,\delta)mMeth}$ distances (to the point C$_{(\alpha,\delta)mMeth}$) of the methanol cloudlets observed by \citet{Bartkiewicz14}, shown at the top. 
The rhombus indicates the V$_{pk}$-V$_{sys}$ velocity against d$_{(\alpha,\delta)mMeth}$ of the \citet{Pandian11} spot that lies in this FOV.
Crosses ($\times$) indicate the locations of the 
\citet{Bartkiewicz14} methanol cloudlets and  
the rhombus, as well as the \citet{Pandian11} methanol cloudlet,
all of them given as offsets in asec, with respect the OH VLBA reference coordinates, shown at the bottom. 
The size of the symbols is proportional to the velocity.
The cross (+) indicates the location where the maximum correlation between V$_{pk}$-V$_{sys}$ and the distance d$_{(\alpha,\delta)mMeth}$ is attained.
The distance, d$_{(\alpha,\delta)mMeth}$, used in the top-panel plot is computed from this point. Dashed circles are plotted for reference at 30 and 130 mas from C$_{(\alpha,\delta)mMeth}$.}
\label{FMetaVelVSdist} 
\end{figure}

The \citet{Pandian11} methanol maser observations also were made to different pointing positions, whose locations span to more than 1$^\prime$ in both, $\alpha$ and $\delta$. 
The locations of the spots are given in absolute coordinates in their Table 3. 
Only the G43.16+0.02 maser spot, lies in the FOV of the OH masers.  
Its location coincides with the above-mentioned \citet{Bartkiewicz14} group (bottom panel of Figure~\ref{FMetaVelVSdist}). 

To transform these coordinates into offsets, with respect to the OH reference point, we subtracted the OH reference coordinates from the \citet{Pandian11} coordinates of the spot. 

The observations of \citet{Bartkiewicz14} were made with a very high angular resolution (of $\sim$5 mas).
Among the parameters that they report, here we use the above-mentioned offsets (RAoffset, Decoffset), the velocity of the peak flux (V$_{pk}$) and the velocity gradient (V$_{grad}$), all of them for each spot. 
The last parameter is the velocity variation at the different velocity channels, where the spot is detected, with respect to the coordinates at these velocities. 

In Figure~\ref{FCoordsROIs}, the locations of the methanol maser cloudlets are represented with crosses.
The spots of both, \citet{Bartkiewicz14} and \citet{Pandian11} data are clumped together in the west, near the large dashed circle.

The OH VLBA field is 6"x6", which is small compared to the whole area of the \citet{Bartkiewicz14} spots at W49N and of the field of the \citet{Pandian11} one.
This is to say that even if the distance between the centres of the fields is approx 5", the methanol fields are much larger than the OH one and many of the methanol cloudlets, as reported by \citet{Bartkiewicz14} and by \citet{Pandian11} lie outside the field of the OH spots.

The peak velocities, V$_{pk}$, of the \citet{Bartkiewicz14} maser spots spread from about 8 to 20 $km~s^{-1}$.
The range of velocities and the high spatial resolution allows us to do the same process as that for OH spots.
A mesh of points, separated $\Delta\alpha$=10 mas, from -4500 to -5100 mas in right ascension (RA) and separated $\Delta\delta$=10 mas from -600 to 0 mas in Declination (Dec) was used.
The correlation and the $\chi^2$ between V$_{pk}$ and distance are estimated to each point of the mesh for the 12 spots mentioned by \citet{Bartkiewicz14}.
We further refer to C$_{(\alpha,\delta)mMeth}$ as the point for which the highest correlation is attained.
In the top panel of Figure~\ref{FMetaVelVSdist}, we show a plot of V$_{pk}$-V$_{sys}$, with V$_{sys}$=8 $km~s^{-1}$, where the systemic velocity estimated by \citet{Welch87} against the distance (d$_{ccMeth}$) to C$_{(\alpha,\delta)mMeth}$ is shown.
It may be seen that the velocity undergoes a gradient with a negative slope (as in the case of OH masers).
Similarly to the procedure with OH data, a straight line with velocity V$_{ftd}$ is fitted to these data.
The V$_{pk}-$V$_{sys}$ against distance to C$_{(\alpha,\delta)mMeth}$ values for the Pandian spot are also plotted with a rhombus.

In the bottom panel of Figure~\ref{FMetaVelVSdist}, the locations of the methanol maser cloudlets of the G43.165+00.013 group from \citet{Bartkiewicz14} are indicated by crosses ($\times$).
The sizes of the symbols are proportional to the V$_{pk}$-V$_{sys}$ velocities.
The rhombus in the N corresponds to the \citet{Pandian11} spot.
The point C$_{(\alpha,\delta)mMeth}$ is indicated with the plus symbol (+).
The dashed circles are plotted for reference at distances from C$_{(\alpha,\delta)mMeth}$ (30 and 130 mas), where more spots are seen.
It may be seen from the bottom panel of Figure~\ref{FMetaVelVSdist} that the spots closer to C$_{(\alpha,\delta)mMeth}$ have higher velocities than the spots farther from it. 
Additionally, it is seen that spots located at different sectors of the small dashed circle have similar V$_{pk}$-V$_{sys}$ velocities.
The velocity field indicates acceleration towards C$_{(\alpha,\delta)mMeth}$.

\begin{figure}[ht!]
\centering
\includegraphics[height=6cm,width=10cm]{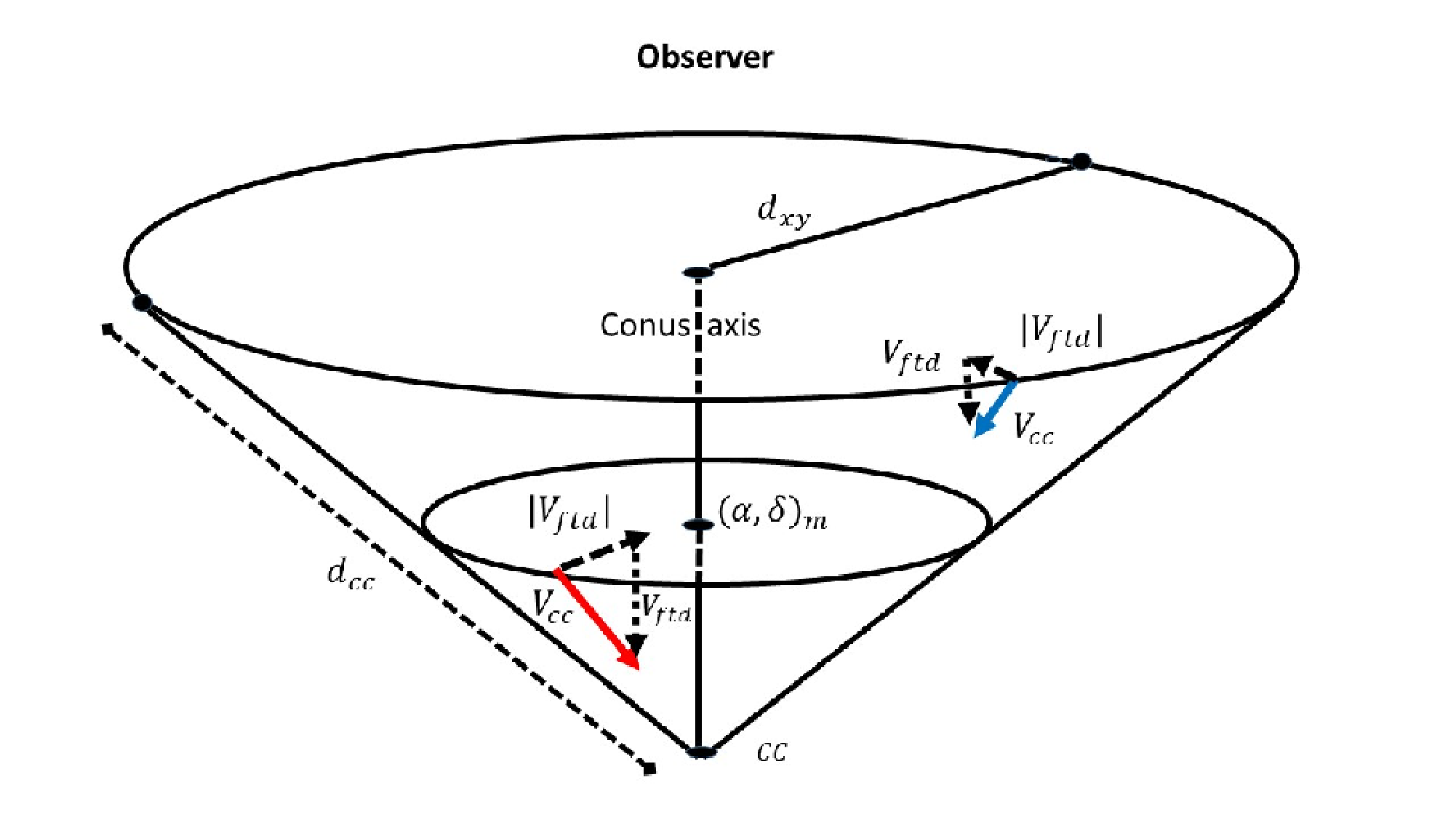}
\caption{Model of the contraction velocities, V$_{cc}$, of the OH spots, i.e. of their velocities towards a common centre of contraction ($CC$).
V$_{ftd}$ (represented with dotted lines) corresponds to the velocity from the straight line fitted to the observed data for the given distance to $(\alpha,\delta)_{m}$, as may be seen in Figure~\ref{FVelvsDistini}.
This velocity is assumed to be the radial component of the contraction velocity.
Each spot has its own velocity (the two V$_{ftd}$, represented here, are different to each other in magnitude).
The small ellipse corresponds to the small circle plotted in Figure~\ref{FCoordsROIs}.
For an assumed spot at a point in this circle, |V$_{ftd}$| is the magnitude of the transversal velocity, which is directed to the CONUS axis (and orientated perpendicular to it). 
The |V$_{ftd}$| of the spot at the large ellipse is smaller than that at the small one.
The resulting contraction velocity, of a magnitude, V$_{cc}$, lies on the surface of the CONUS and it is directed to the $CC$ point (the apex of the CONUS).}
\label{FmodelConus}
\end{figure}

\pagebreak

\section{Discussion}
The velocity field in 3D could be reconstructed by using both
the radial velocity, V$_{obs}$, and the proper motions of the spots.
However, in this case, we did not have access to this last information. 
Nevertheless, the fact that there is a region, 
in the field, for which the correlation between 
V$_{obs}$ and the distance to that region,
d$_{(\alpha,\delta)m}$, takes the 
largest values, among all the points of the field; this
indicates that the behaviour 
of the velocity field is more clearly revealed
because it is probably a 
location 
the behaviour of this velocity field is dependent on.
In other words, it is a parameter involved in a functional
relationship between V$_{obs}$ and d$_{(\alpha,\delta)m}$.

For Keplerian rotation, the velocities in the E and W sectors,
with respect to the systemic velocity,
are opposite to each other and the velocities in the N and S are near to the value of the systemic velocity, as can be seen at single sources. 
From Figures~\ref{FVelvsDistini} and
\ref{FCoordsROIs}, it may be seen
that none of these characteristics is present 
in the OH spots in W49N. 
Nevertheless, a series of tests of the locations and velocities of 
the sources was made with a geometric method, which allows us
to compare them with ellipses, which are due to inclined 
Keplerian rotation orbits \citep{Rodriguez11}.
In Appendix~\ref{appendix A}, the bases of the method to test the fit of the spots to an ellipse are given.
Applying it to OH maser data, it is found that only nine spots of the 1665 LCP and ten spots of 
the 1665 RCP provide a good fit to ellipses for Keplerian-rotation.
However, the geometric elements of the ellipses
(centre, major axism and eccentricity)  differ 
considerably from each other.
Finally, we conclude that the spots are not in a rotating disk.

\begin{figure}[ht!]
\centering
\includegraphics[height=6.3cm,width=6.6cm]{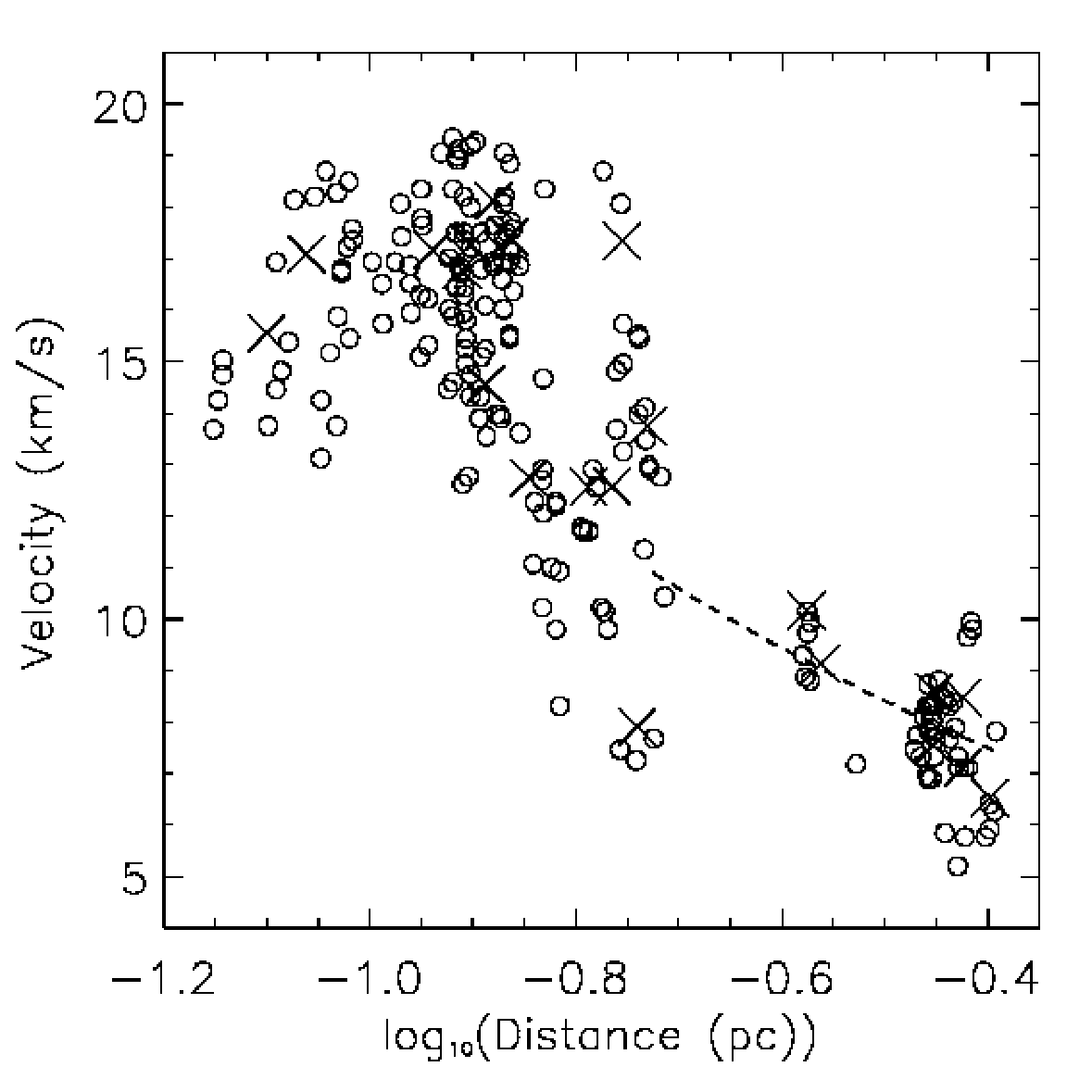}
\caption{Estimated total velocity, V$_{Tot}$, of the OH spots to the centre of contraction, 
as a function of the distance to $CC$, d$_{cc}$.
The crosses ($\times$) denote the average velocities for the groups
indicated with bars in Figure~\ref{FCoordsROIs}.
The dashed line denotes the free fall velocity for M$_{inn}$=2500 M$_{\odot}$.}
\label{FVelVSdistPCcc}
\end{figure}

\begin{figure}[ht!]
\centering
\includegraphics[height=6.3cm,width=6.6cm]{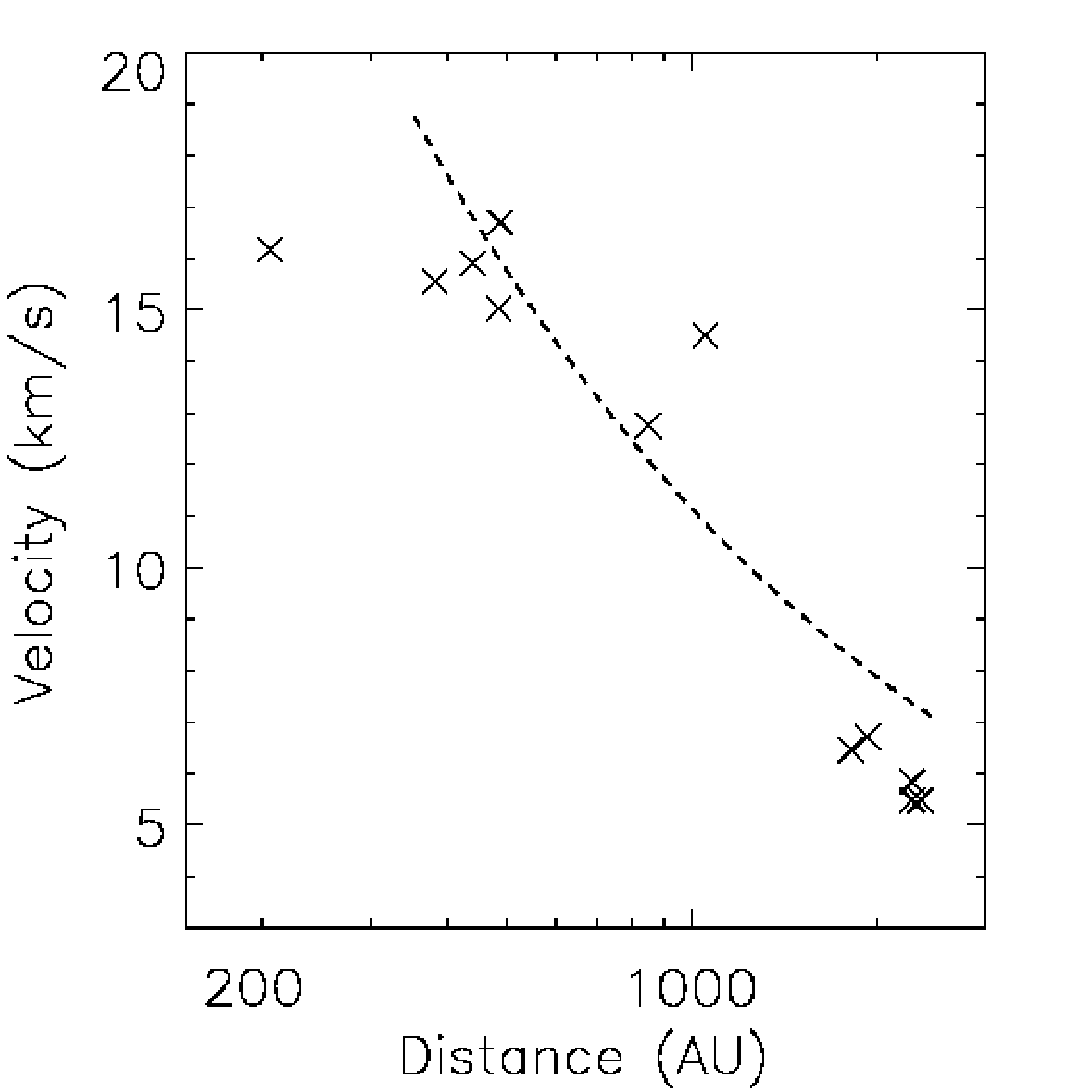}
\includegraphics[height=6.3cm,width=6.6cm]{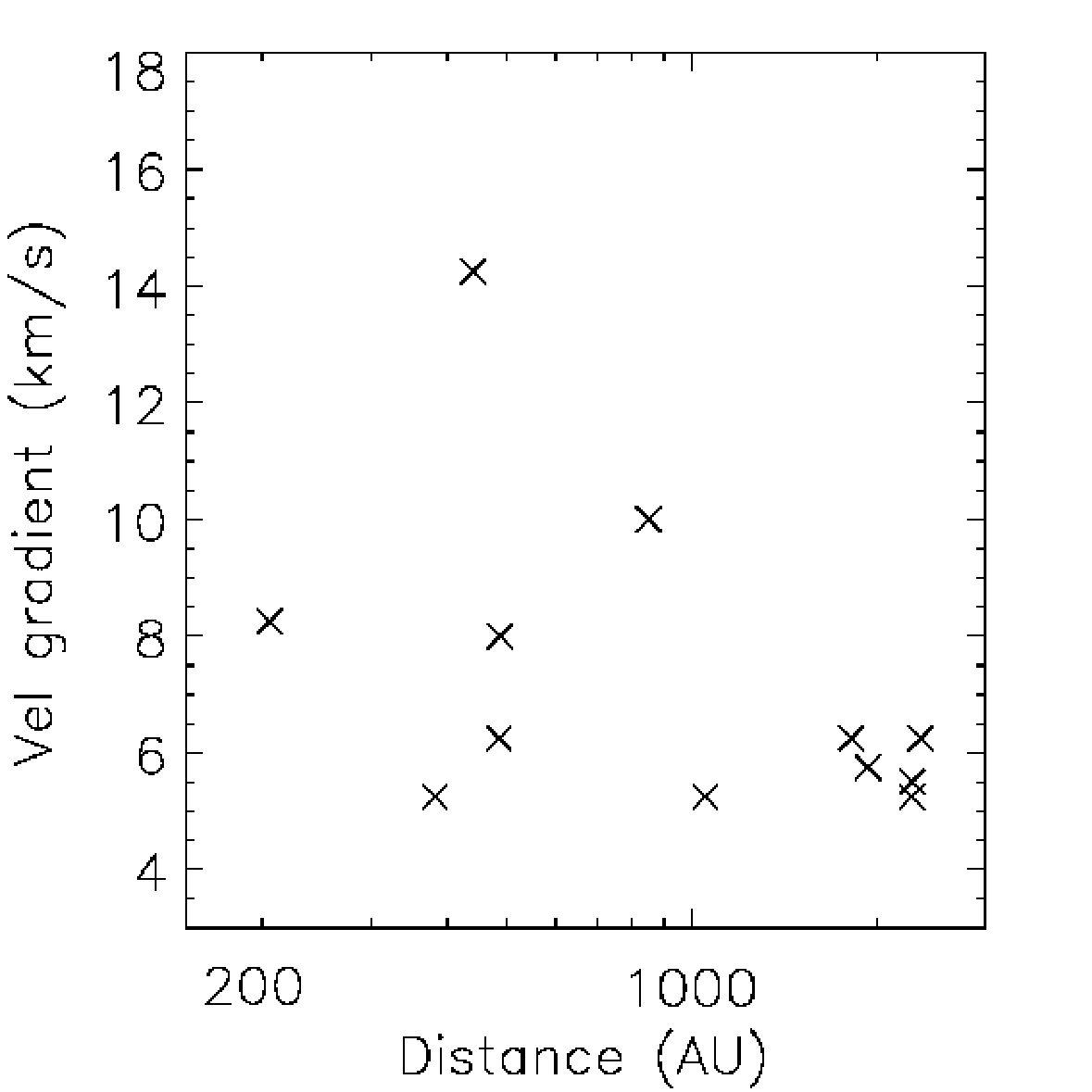}
\caption{Velocity V$_{TMeth}$ versus the distance, of each methanol cloudlet seen at the $W$ in Figure~\ref{FCoordsROIs}, to d$_{ccMeth}$ shown on top.
The dashed line corresponds to the free-fall velocity for a 75 M$_{\odot}$ inner mass. 
{\it } The V$_{grad}$ velocity of each methanol cloudlet versus the d$_{CCMeth}$ distance (bottom).}
\label{FBartCoordyVccDcc} 
\end{figure}

It has been seen that V$_{obs}$ shows a general trend 
along the whole field, indicating that the velocity 
of the masers, in this field, has a global component. 
With respect the $(\alpha,\delta)_m$ point, the
spots are predominantly in the $E$ sector,
distributed in an elongated shape, $NE-WS$ orientated.
In other sources, accretion has been seen to take 
place through filaments \citep{Peretto13}. 
However, the amount of material accreted through 
them is not as large as the amount accreted radially.
In the case of the W49N, the highest correlation
is observed between V$_{obs}$ and d$_{ij}$
(the distance to the point $(\alpha,\delta)_m$) and 
not along the direction of the elongation.
The OH spots with similar velocities are approximately at the same 
distance to $(\alpha,\delta)_m$. 
For example, as above mentioned, the spots that are close to the 
reference circle plotted at 1.6 asec radius have similar velocities,
although they are in different sectors. 
For the spots around the circle at 4.6 asec, the same situation is seen but these spots have lower velocities.
The spots are likely in a global velocity field,
which is consistent with an accelerating contraction 
towards the $(\alpha,\delta)_m$ point; this can be interpreted as global contraction towards a common centre (CC).
The results obtained with VLBA observations are supported by the \citet{Argon03} data, obtained with the VLA, the $(\alpha, \delta)_m$, as these data only differ by 0.3 asec with respect to the point 
estimated with the VLBA data.

\subsection{3D model of the contraction velocities} 
\label{SecModelCC}
We recall that V$_{ftd}$ has a functional relation (fitted line in Figure~\ref{FVelvsDistini}) to the distance to $(\alpha,\delta)_m$.
The shorter the distance to $(\alpha,\delta)_m$ the greater the V$_{ftd}$.
Then, given the distance to $(\alpha,\delta)_m$, the corresponding V$_{ftd}$ velocity can be
known.
It is the same for all the points on a circle centred at $(\alpha,\delta)_m$.
These velocities can then be further used for the purposes of building a model.
In Figure~\ref{FmodelConus}, the V$_{ftd}$ velocity for two spots, one assumed to be at the distance of the small and the other of the large circles of Figure~\ref{FCoordsROIs} are shown.
Each V$_{ftd}$ is considered to be the radial component of the contraction velocity to a common centre (V$_{cc}$).
They differ from one  another in magnitude and, ultimately, that the spot that is closer to $CC$ has higher V$_{ftd}$ velocity than the spot that is farther from it.
We consider the possibility that the $CC$ is not on the same plane of the small circle (at Figure~\ref{FCoordsROIs}).
We assume that the V$_{cc}$ of the two plotted spots and of all the spots are directed towards $CC$, which is at the apex of a CONUS, and all the V$_{cc}$ vectors are located on the surface of the CONUS.

In the model, $d_{xy}$ is the distance from a spot to the CONUS axis (Figure~\ref{FmodelConus}).
In particular, for the plane of the small circle, this distance is equal to $d_{(\alpha,\delta)m}$. 
Furthermore, the XY-plane of the small circle will be considered as a reference.
It is represented with the small ellipse in Figure~\ref{FmodelConus}. 
The distance to this plane is $d_z$, it is positive as going farther from the observer and negative as it gets  closer to the observer.
In terms of the observed values, it may be seen that
the spots that are at distances < 1.6 asec, 
that is, closer to $CC$ with respect to the small circle and farther from the observer
would be at $d_z$>0; the spots 
that are at distances >1.6 asec, namely,  farther from $CC$ and closer to the observer would be at $d_z$<0 (Figure~\ref{FmodelConus}).
The $CC$ is in the line of sight of the point $(\alpha, \delta)_m$
at a distance, d$_{z0}$.
The angle between the CONUS axis and the line 
from a spot to the $CC$ is indicated with $\theta$.

The tangential component of V$_{cc}$ is unknown.
Then, for the model, the following assumptions are as follows:
1) the tangential component of V$_{cc}$ follows the same functional relation with the distance as the radial component with the distance to $(\alpha, \delta)_m$.
We remember that in the z-direction, the distance to $CC$ is d$_z$ (Figure~\ref{FmodelConus});
2) it is orientated perpendicular to the CONUS axis and it directed to it.
In the case of spots on the small circle of Figure~\ref{FmodelConus}, this velocity is pointing to $(\alpha,\delta)_{m}$.
3. It can be considered, for simplicity, that the tangential component of V$_{cc}$ has a magnitude equal to $|V_{ftd}|$.
In the representation of Figure~\ref{FmodelConus}, each point on the CONUS has its own velocity V$_{ftd}~(\hat{z})$, pointing in a radial direction (respect the observer) and their own tangential velocity,|V$_{ftd}|(\hat{r})$ (pointing to the CONUS axis), and the vector $\overrightarrow{V}_{cc}$ would be the sum of them.
Furthermore, these velocities are not written in vectorial notation, since in the model, each of them displays a clear orientation and sense, as may be seen from Figure~\ref{FmodelConus}.

With the above-listed conditions on the contraction velocities, we can assume that the distance of a spot to the CONUS axis (d$_{xy}$), which is measured perpendicular to the axis, would be the same as the distance in the radial direction (d$_{z}$) and thus $\theta$=45$^{\circ}$.
This means that for the plotted spots (and any spot overall even if it does not lie at any of the two circles), the result will be d$_{z}$=d$_{xy}$.
The distance of each spot to $CC$ is thus d$_{cc}$ =$\sqrt{2}$~d$_{xy}$.
The distance, d$_{cc}$, is measured from $CC$ to the spot. 
In the model (Figure~\ref{FmodelConus}), the distance, d$_{cc}$, to a spot that lies within the dashed circle, namely, at a distance of 1.6 asec, is indicated.
If another spot does not lie at this distance, then it will not be at the dashed circle; however, it will be over the CONUS and its infalling velocity (due to global contraction) that is given for the linear fit of Figure~ \ref{FVelvsDistini}.
The tangential velocity also depends on $\theta$, which is the aperture of the CONUS.
For $\theta$=45$^{\circ}$, its magnitude is $|V_{ftd}|$ and then V$_{cc}$=$\sqrt{2}$~V$_{ftd}$.

Besides the V$_{cc}$ values, the velocity of each spot also could have random or local kinematics components.
However, the difference $\Delta$V=(V$_{obs}$-V$_{sys}$)-V$_{ftd}$ (Figure~\ref{FVelvsDistini}) represents only the radial component of this velocity.
To include a reasonable input to the random or local kinematics velocity in a tangential direction, we compute it in the same way as for $V_{ftd}$.
Then, taking into account that $\theta$ = 45 $^{\circ}$, the estimation of a total velocity for each spot leads to $V_{tot}$ = V$_{cc}$ +$\sqrt{2}$~$\Delta$V.
The total velocity is plotted in Figure~\ref{FVelVSdistPCcc}.


\begin{figure}[ht!]
\centering
\includegraphics[height=6.3cm,width=6.6cm]{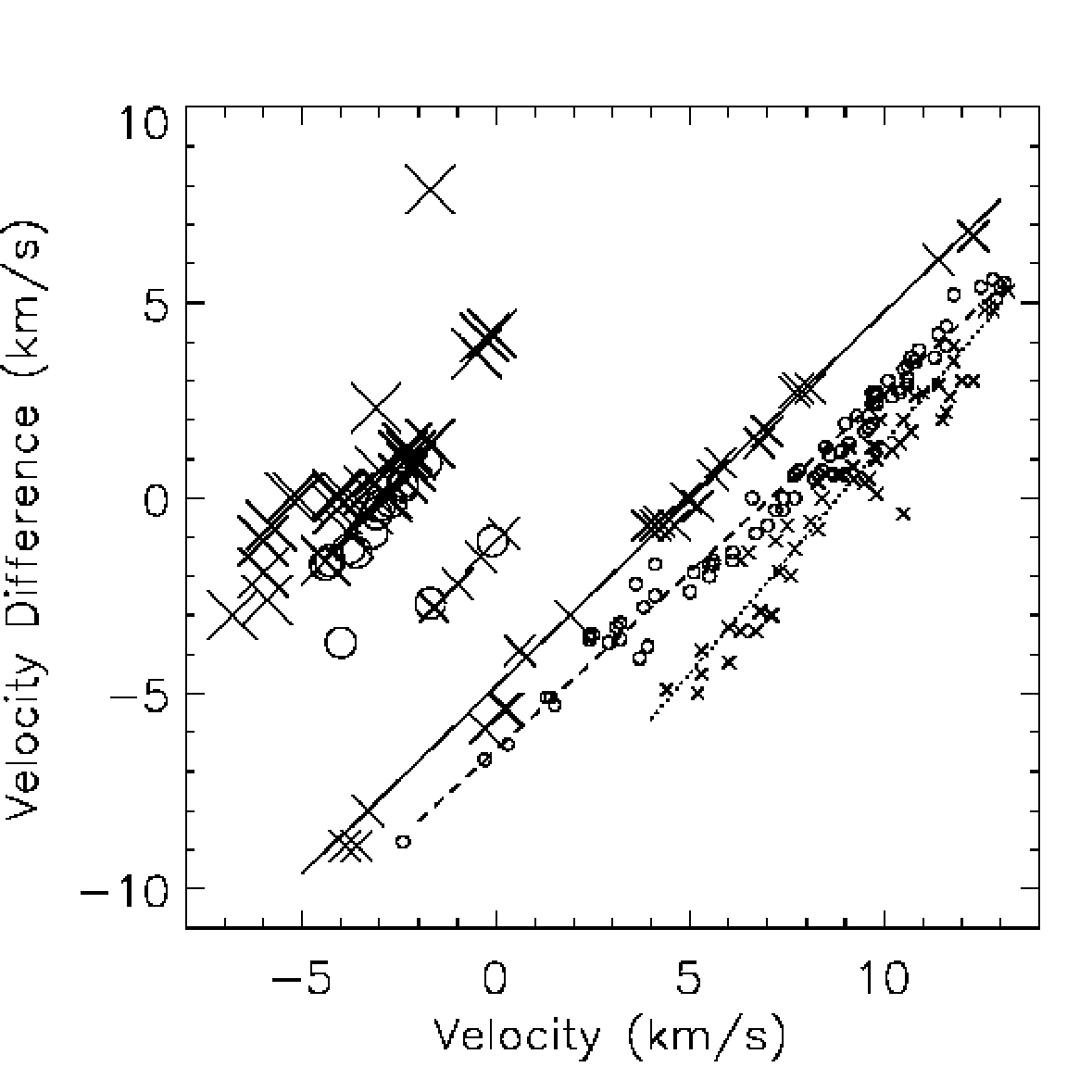}
\caption{Velocity difference ($\Delta$V) versus V$_{obs}$.
Different symbols correspond to spots at different distance
intervals.
The small crosses are for d$_{cc}$< 0.09 pc,
the dashed line is fitted to these spots. 
The small circles are for spots at 0.09 pc <d$_{cc}$< 0.12 pc,
and the dotted line is fitted to them.
The continuous line is fitted to spots at 0.12 pc <d$_{cc}$< 0.19 pc, 
which is shown with mid-size crosses.
The large circles are for 0.19 pc <d$_{cc}$< 0.25 pc and
large crosses for 0.25 pc <d$_{cc}$< 0.42 pc.}
\label{FDifVelVSvel}
\end{figure}

The average velocity of groups of spots located near each other is an estimate of the group velocity and this would reduce the impact of local kinematics or random velocities in the data dispersion. 
This could allow us to see how the group velocity versus d$_{cc}$ would behave, with regard to the case of separated spots. 
The groups of spots that we used are indicated with the vertical lines in Figure~\ref{FCoordsROIs}. 
The average of the velocity of each group is shown with
diamonds in Figure~\ref{FVelVSdistPCcc},
together with the velocity of separated spots (crosses).
It may be seen that they behave similarly to each other, showing a maximum at about log(d$_{cc}$)=-0.9$ (\sim$ 0.12 pc).
A general velocity increase is seen, which is rather noisy nonetheless, 
as a trend from 0.19 to 0.12 pc.
On the other hand, a general decrease from 0.12 pc to 
shorter distances is also seen.
For the computation of the infall velocity field, we considered
the 0.42-0.19 pc interval.
For the infall velocity, we use the value fitted to
V$\sim d_{cc}^{-0.5}$ at 0.19 pc, which is 10.7 $km~s^{-1}$.

We recall that the velocity difference ($\Delta$V) is the result of subtracting the fitted velocity, given by the straight line of Figure~\ref{FVelvsDistini} from the observed velocity.
Therefore, $\Delta$V is related to components, of the spot velocity, which is different to the infalling velocity.
In Figure~\ref{FDifVelVSvel}, the velocity difference, 
$\Delta$V versus V$_{obs}$ is plotted.
The closer a spot is to $CC$, the smaller the symbol that represents its velocity. 
Different symbols correspond to spots at different distance intervals.
The spots grouped in a given distance interval are fitted by the same straight line.
This occurs even though the spots are located at 
different sectors with respect to a centre point, which in this case, is $CC$.
The small crosses represent the closest spots to $(\alpha,\delta)_m$,
while small circles represent the spots that are next in terms of the distance and so on.
The spots around the continuous line 
correspond to the velocity increase seen from 0.19 to 0.12 pc. 
As may be seen from this figure, these lines show the largest velocity dispersion.
They belong to different groups of Figure~\ref{FCoordsROIs},
some of them located in the east, others in the south, 
and some in the north-west. 
The spots at d$<$0.12 pc also belong to various groups
of Figure~\ref{FCoordsROIs} in different sectors. 
They seem to be experiencing a deceleration as approach
CC (Figure~\ref{FVelVSdistPCcc}).
This could be similar to the case of 
a single protostar model \citep{Stahler80}
in the sense that it could be accretion in a structured
shape, including different layers.
In the case of global collapse, there is no such detailed 
scheme.
However, here we do not attempt to study these details in this work.

To estimate the free fall time, we assumed 
n=1$\times 10^{-4}$ cm$^{-3}$ and an average 
mass particle $<$m$>$=2.33 m$_H$ 
\citep{Kauffmann08, Kauffmann13}, 
obtaining a result of t$_{ff}$= 3.4$\times 10^5$ yr.
The accretion rate, assuming a spherical geometry, can be 
calculated from $2\pi r \rho V_{infall}$, for a distance 
of 0.19 pc and $\rho$ from the above estimation,
we have an accretion rate of 1.4$\times 10^{-3}$ M$_{\odot}yr^{-1}$. 
Considering this accretion rate, the time required to accumulate 
a mass of 2500 $M_{\odot}$ is 1.8$\times 10^6$ years, which is 
5.2 times t$_{ff}$. 

Part of the inner mass is in the cores and stars embedded
in HII regions.
\citet{Smith09} found that several radio sources are deeply
embedded inside the cloud core.
Also, part of the mass could take the form of stars
that already have swept out their HII regions and could
be not detectable in the radio continuum and other wavelengths.

To estimate the virial coefficient, it must be taken 
into account that the velocity dispersion is due 
to global and to local kinematics. 
We calculated the virial mass, M$_{vir}$, using the standard deviation
of $\Delta$V, which turns out to be 2.9 $km~s^{-1}$
and leads to M$_{vir}$=380 $M_{\odot}$. 
It is smaller in more than six times M$_{inn}$,
which is the mass obtained by fitting the free-fall function to 
the V$_{cc}$ values. 

\citet{Rudolph90} estimated a global collapse in W51 with 
a mass of 4$\times 10^4$ $M_{\odot}$ and a 1200 M$_{\odot}$
subcollapse. 
The field of the OH spots in W49N has a side of
approximately $\sim$7 asec, while the ring of \citet{Welch85} for W49N is $\sim$ 20 asec.
This means that the smaller region is virialized by itself, even though
it is a fraction $\sim$(7/20)$^3$ of the Welch ring of HII regions.
However, the kinematics of the smaller region spans a
velocity range similar to the velocity range of the ring.
The global velocity field of OH masers could be due to a 
sub-collapse similar to that observed in W51.

The analysed group of methanol cloudlets show similar conditions, in velocity and location, to those seen in the OH masers, allowing us to apply them to the same model (with another centre of contraction).
It should be noted, however, that only one OH maser spot (indicated with a circle in the right panel of Figure~\ref{FMetaVelVSdist}) is co-spatial to the methanol cloudlets.
This means that the OH spots do not trace this region.
\citet{Sanna10b} studied the high-mass star-forming region G23.01-0.41 and found that the OH spots are more distant from the YSO than the methanol ones.
Here, the only OH spot, which is co-spatial to this group, is certainly in the external region with respect to CC$_{Meth}$.

The velocity gradient, seen at a group of methanol cloudlets, could be related to the kinematics of the source.
In the top panel of Figure~\ref{FBartCoordyVccDcc}, the velocity gradient, V$_{grad}$, of each spot is plotted against its distance to CC$_{Meth}$.
It may be seen that the values of V$_{grad}$ at $\sim$500 AU (near to the small dashed circle) have a larger dispersion than the spots at $\sim$2000 AU (near to the large dashed circle).
We should not exclude the possibility that near to CC$_{Meth}$, another kinematical component, such as a disk, jet, or outflow that has been traced in other sources with methanol masers \citep{Pandian11, Moscadelli11}, could be present.
That scenario would lead to larger dispersion in the velocity field in this region and a larger difference in the velocity from one spot to another than in regions that undergo a single type of kinematics.

Only one \citet{Pandian11} methanol maser cloudlet lies in the field of OH spots, which is coincident with the \citet{Bartkiewicz14} methanol cloudlets.
They are located in the west of the VLBA field (crosses ($\times$)
The velocity of the \citet{Pandian11}
spot is V$_{pk}$=9.3 $km~s^{-1}$ which, taking into account V$_{sys}$=8 $km~s^{-1}$, gives us V$_{pk}$-V$_{sys}$=1.3 $km~s^{-1}$.
This velocity is similar to those of the OH spots located at similar distances from CC$_{OH}$.
The values (velocity and distance) for this spot are also plotted in Figure~\ref{FMetaVelVSdist}. 
The rhombus that represents it (in the top panel) lies near the straight line fitted to the velocity versus distance plot, for the \citet{Bartkiewicz14} spots.

\subsection{Contraction of the group of methanol maser cloudlets} 
Velocity gradients are expected to be observed in different kinematical models.
In the case of maser spots in a rotating disk, the spots with more red- and more blue-shifted velocities would be observed, from the Earth, at both sides of the central region.
The locations at W49N of methanol cloudlets with similar velocities, in different sectors (small dashed-circle of the right panel of Figure~\ref{FMetaVelVSdist}), differ from what is expected in a rotating disk.
Instead, this indicates that another type of kinematics is responsible for spots with such velocities and locations.
The expansion or contraction is suitable to explain both conditions, with respect to a given centre. 
A third condition, namely, the larger velocities seen closer to the C$_{(\alpha,\delta)mMeth}$ point, agree with accretion in an accelerated velocity field.
 
The velocity versus distance relation of the methanol maser cloudlets seen at the FOV of the OH VLBA maser spots (Figure~ \ref{FMetaVelVSdist}) also suggests a centre of contraction for this particular group.
A model of contraction is applied in a similar way as in the case of OH spots (Section~ \ref{SecModelCC} and Figure~ \ref{FmodelConus}).
We assume that the distance in the direction perpendicular to the XY-plane, from each spot to CC$_{Meth}$, is the same as the distance, d$_{(\alpha, \delta)Meth}$, from the spot to $C_{(\alpha,\delta)mMeth}$.
Then, the distance of each spot to this point is d$_{ccMeth}$ =$\sqrt{2}$~d$_{(\alpha, \delta)Meth}$
and we consider that the total velocity is V$_{TMeth}=\sqrt{2}$V$_{ftd}$+$\sqrt{2}\Delta V_{Meth}$.
The first term corresponds to the falling velocity (towards $CC_{Meth}$) and the second to the dispersion velocity, with respect to the falling one.

In the top panel of Figure~\ref{FMetaVelVSdist}, the velocity, V$_{TMeth}$, is plotted against d$_{ccMeth}$.
Taking the interval of distances from 300 to 2500 AU, a free-fall function is fitted to V$_{pk}$-V$_{sys}$ versus d$_{ccMeth}$ (dashed line in the top panel of Figure~\ref{FBartCoordyVccDcc}), obtaining an inner mass of 75  M$_{\odot}$.
This approximation helps us to estimate the inner mass, under the assumption of accretion.
In this case, it is a region on a smaller spatial scale than that of the OH maser spots,
indicating the possibility of another sub-collapse (that could contain various forming stars or one massive forming star).

The amount of methanol cloudlets in the W49N field, where the OH masers are seen, is smaller than the OH spots number.
The methanol cloudlets are at given locations of the whole region, while the OH ones are more spread, namely, they appear in more locations across the region.
This result agrees with the previous finding that 6.7 GHz methanol masers trace early phases of the high-mass star formation process \citep{Minier05, Pandian11}.
The methanol masers could be more suitable to trace the velocity of the environment of such objects, providing the opportunity to study their kinematics.
This also indicates that the OH and methanol masers seem to be complementary to each other in the study of star formation rates 
(SFR, \citep{Sanna10b}).

The locations and velocities of methanol masers do not disagree with a scenario of a subcollapse in W49N, as deduced from OH masers.
Moreover, their velocities and locations agree with accretion scenarios at shorter spatial scales, which could indicate the possibility of sub-collapses on these scales.
The combination of both maser species, therefore, seems to be a tool that is useful in building a more complete picture of the kinematics of SFR on different spatial scales, particularly in large complexes such as W49N.

As opposed to the case of the OH and the methanol maser spots analysed here, which trace contractions, the H$_2$O maser spots at W49N, trace outflows.
The bipolar outflow is related to the local kinematics at the G1 HII region and the velocity range of the H$_2$O maser spots is considerably larger than the global field velocity range estimated by \citet{Welch85} and also for the one  analysed here.
This means the H$_{2}$O spots are related to local conditions at a given HII region, while the OH maser spots are related to the global kinematics.

\section{Conclusions}
The maximum correlation between the observed velocity,
V$_{obs}$, and the distances to test points, d$_{ij}$, 
takes values larger than 0.7 for 1665 (RCP and LCP) and
1667 (RCP and LCP).
Using the distance of the spots to the weighted average
coordinates $(\alpha, \delta)_m$, of the maxima of the 
correlation, the location of the projection of the centre of contraction to the XY-plane is estimated as ($\alpha_{2000}$=19:10:13.1253, $\delta_{2000}$=9:6:13.570).
It is found that the behaviour of 
V$_{Tot}$=V$_{cc}$+$\sqrt{2}$ $\Delta$V against d$_{cc}$ 
shows a maximum at 0.12 pc, with a decrease from 
0.12 to 0.19 pc that takes place faster than the
decrease between 0.19 and 0.42 pc.
Assuming that the behaviour of the velocity at the 0.19 
and 0.42 pc interval is due to free fall accretion, 
an inner mass of M$_{inn}$=2500 $M_{\odot}$ is obtained.
We estimate the accretion rate to be 
$\dot M$=1.4$\times$ 10$^{-3}$ M$_{\odot}$yr$^{-1}$ 
which requires a time t$_{inn}$=1.8$\times$10$^6$ yr 
to accumulate M$_{inn}$.
The velocities of the OH spots at W49N, together with their 
positions with respect to the $(\alpha, \delta)_m$ point, makes it possible 
to estimate the spot kinematics. Finally, this can be interpreted as 
a contraction of the cloud that appears to be a sub-collapse 
in the W49N MC.

\section{Acknowledgements}
We would like to thank the Astrophysics Department of INAOE
for its financial support. M-T J.E. would like to acknowledge M. Goss 
for providing the VLBA data of 2005.
We also thank the anonymous referee for her/his careful review of
the present article and helpful suggestions, which have been very useful to us for improving this work.

{}

\begin{appendix}
\section{Model of ellipse} 
\label{appendix A}
To test if the observed positions of the maser spots fit into a model of an ellipse, the equation of the conic is used, given as follows:

\begin{centering}
\begin{equation}
G=a_{11} x_{i}^{2}+ a_{12} x_{i} y_{i}+a_{22} y_{i}^{2}+ a_{01} x_{i}+ a_{02} y_{i}+ a_{00}
\label{EGeqa11a22}
,\end{equation} 
\end{centering}

\noindent where $x_{i}$ and $y_{i}$ are, respectively, the observed RA and observed Dec. of the spots. 
The coefficients $a_{ij}$ determine the output parameters of the geometric model.
The second-degree coefficients $a_{11}$, $a_{22}$, and $a_{12}$, determine whether the function is an ellipse or not,
$a_{12}$ is the rotation coefficient, while $a_{01}$ and $a_{02}$ are the translation coefficients (Apostol 2001).
Furthermore, we provide a basis for testing whether the spot locations can fit an ellipse, based on Equation~\ref{EGeqa11a22}.

In the teotherical case, for a non-degenerate conic, G=0.
In an experimental situation, the better a model fits the observed values ($x_{i}$ and $y_{i}$), the smaller will be G.
Then, the coefficients have to be selected such that G takes the smallest possible value.
To find the $a_{ij}$ values that
lead to the best fit, the least-squares method is used.
With this purpose, the python function $leastsq$ is applied, from the $optimize$ package inside the $scipy$ python module and $pylab$ interface for the use of the $matplotlib$ library.
To find the coefficient for the best fit, at least six spots are needed.

The least-squares method minimizes the value of $S^{2}$, given by: 

\begin{centering}
\begin{equation}\label{ec.m1}
S^{2}=\frac{1}{N-Np}\sum_{i}^{N}\frac{G^{2}}{ \epsilon_{i}^{2}}
,\end{equation} 
\end{centering}

\noindent where $\epsilon_{i}$ is the estimated uncertainty given as
$\epsilon_{i}^2=\epsilon_{ix}^2+\epsilon_{iy}^2$, with $\epsilon_{ix}$
the uncertainty in $x_{i}$ and $\epsilon_{iy}$ the uncertainty in $y_{i}$, Np is the number of parameters for the fit, N is the total amount of spots, and N-Np is the number of degrees of freedom. 

The Equation~\ref{EGeqa11a22} can be expressed in matrix form as: 

\begin{centering}
\begin{equation}\label{matrix}
G=\begin{pmatrix} 1 & x  & y \end{pmatrix} \begin{bmatrix}
a_{00}&\frac{a_{01}}{2} &\frac{a_{02}}{2}\\
\frac{a_{10}}{2}&a_{11}&\frac{a_{12}}{2}\\
\frac{a_{20}}{2}&\frac{a_{21}}{2}&a_{22}\\
\end{bmatrix} \begin{pmatrix} 1 \\ x  \\ y \end{pmatrix}
\end{equation}
\end{centering}

\noindent and the matrix will be further referred to as $A$, which can be expressed as:

\begin{centering}
\begin{equation}
A=\sum_{i=0}^{2}\sum_{j=0}^{2}a_{ij}=\begin{bmatrix}
a_{00}&\frac{a_{01}}{2} &\frac{a_{02}}{2}\\
\frac{a_{10}}{2}&a_{11}&\frac{a_{12}}{2}\\
\frac{a_{20}}{2}&\frac{a_{21}}{2}&a_{22}\\
\end{bmatrix}
\label{matrixA}
.\end{equation}
\end{centering}

The condition for a non-degenerate conic is that $det(A)\neq0$
and the condition for having an ellipse is that $4a_{11}~a_{22}-a_{12}^{2}>0$.
From $A,$ we can take a submatrix $A_0$ as follows 

\begin{centering}
\begin{equation}
A_{0}=\begin{bmatrix}
a_{11}&\frac{a_{12}}{2}\\
\frac{a_{21}}{2}&a_{22}\\
\end{bmatrix}
\label{EqA0matrix}
.\end{equation}
\end{centering}

Then, considering that $a_{12}=a_{21}$, the condition for an ellipse is $det(A_{0})>0$.
Further, a description of the process to find the elements of an ellipse is made.
It is based on Apostol (2001), where similar cases are studied 
\footnote{The involved equations are given in sections "Reduction from a real quadratic form to a diagonal form", "Applications to analytical geometry" and in a number of examples and exercises.}.
The reduced equation of the ellipse (from the general Equation~\ref{EGeqa11a22}), after a rotation and a translation, is: 

\begin{centering}
\begin{equation}\label{reducida}
\lambda_{1}x'^{2}+\lambda_{2}y'^{2}+g_{00}=0
,\end{equation}
\end{centering}

\noindent where $x'$ and $y'$ are the rotated values (from observed RA and Dec., respectively), $\lambda_{1}$ and $\lambda_{2}$ are the eigenvalues of $A_{0}$ (here, $\lambda_{1}$ is the eigenvalue with the lowest absolute value, since it is inversely proportional to $a$, with $a$ as the major semiaxis, and $\lambda_{2}$ is inversely proportional to the minor semiaxis, $b$), and $g_{00}$ is given by:

\begin{centering}
\begin{equation}\label{g00}
g_{00}=\dfrac{det(A)}{det(A_{0})}
.\end{equation}
\end{centering}

\noindent If Equation~\ref{reducida} is divided by $-g_{00}$ we obtain

\begin{centering}
\begin{equation}\label{reducida2}
\dfrac{\lambda_{1}x'^{2}}{-g_{00}}+\dfrac{\lambda_{2}y'^{2}}{-g_{00}}-1=0.
\end{equation}
\end{centering}

\noindent Equation~\ref{reducida2} can be expressed in a convenient form to get it in the standard ellipse formula, as follows:

\begin{centering}
\begin{equation}\label{reducida3}
\dfrac{x'^{2}}{\left( \sqrt{-\dfrac{g_{00}}{\lambda_{1}}}~~\right)^{2}}+\dfrac{y'^{2}}{\left(\sqrt{-\dfrac{g_{00}}{\lambda_{2}}}~~\right)^{2}}=1
.\end{equation}
\end{centering}

\noindent To diagonalize $A_{0}$ and eliminate the rotation terms,  
with the aim to determine the eigen values, $\lambda_{1}$ and $\lambda_{2}$, the following equation is used (Apostol 2001):

\begin{centering}
\begin{equation}
det(\lambda I-A_{0})=\lambda^2-Tr(A_{0})\lambda+det(A_{0}),
\label{EqlambdaIA0}
\end{equation}
\end{centering}

\noindent where $I$ is the identity matrix and $Tr(A_{0})$ is the trace of the matrix $A_{0}$.
The solution of the quadratic Equation~\ref{EqlambdaIA0} yields two values, namely, $\lambda_{1}$ and $\lambda_{2}$. On the other hand, it may be seen from Equation~\ref{reducida3} that the minor and major axis are given in terms of the eigen values as:

\begin{centering}
\begin{equation}\label{ec.a}
a=\sqrt{-\dfrac{g_{00}}{\lambda_{1}}}
\end{equation}
\end{centering}

\noindent and

\begin{centering}
\begin{equation}\label{ec.b}
b=\sqrt{-\dfrac{g_{00}}{\lambda_{2}}}
,\end{equation}
\end{centering}

\noindent where the eigenvalues $\lambda_{1}$ and $\lambda_{2}$ are found from Equation~\ref{EqlambdaIA0} by using the $A_0$ matrix and, consequently, the $a_{ij}$ values that lead to the minimum of $S^{2}$ in Equation~\ref{ec.m1}. These are the values that give the ellipse of the best fit.

The centre of the ellipse, ($x_{0},y_{0}$), can be found from Equation~\ref{EGeqa11a22}.
Substituting the translation equations $x=x'+x_{0}$ and $y=y'+y_{0}$ into Equation~\ref{EGeqa11a22}, and assuming it its equals zero, we have:

\begin{centering}
\begin{eqnarray}
0=a_{11} x_{i}^{'2}+ a_{12} x'_{i} y'_{i}+a_{22} y_{i}^{'2}+(a_{10}+2a_{11}x_{0} +a_{12}y_{0}) x'_{i} \nonumber \\+ (a_{20}+a_{21}x_{0}  +2a_{22}y_{0}) y'_{i}+ g_{00}
.\end{eqnarray}
\end{centering}

\noindent Then, setting the terms related to the translation  
equal to zero, that is, the coefficients of $x'$ and $y'$, it follows that

\begin{centering}
\begin{equation}\label{centro1}
a_{10}+2a_{11}x_{0} +a_{12}y_{0}  =0
\end{equation} 
\end{centering}

\noindent and

\begin{centering}
\begin{equation}\label{centro2}
a_{20}+a_{21}x_{0}  +2a_{22}y_{0}  =0
.\end{equation}
\end{centering}

\noindent The values of $x_{0}$ and $y_{0}$ can be found 
by using the values of the model of the ellipse, which lead to the best fit, then solving the system of Equations~ \ref{centro1} and \ref{centro2}, which correspond, respectively, to the second and third rows of the matrix $A$.

Finally, the inclination angle ($\theta$) is given as (Wooton 1985):

\begin{centering}
\begin{equation}
tg(2\theta)=\frac{a_{12}}{a_{11}-a_{22}}
,\end{equation}
\end{centering}

\noindent which can be found by using the values for the best fit.

\end{appendix}

\end{document}